\documentclass[review,authoryear]{article}
\usepackage{graphicx}
\usepackage{url} 
\usepackage{amsmath}
\usepackage[ruled]{algorithm2e}
\usepackage{subcaption}
\usepackage[noend]{algpseudocode}
\usepackage{color} 
\DeclareMathOperator{\diag}{diag}
\usepackage{bm}
\usepackage{natbib}
\usepackage[colorlinks,citecolor=blue,linkcolor=blue,urlcolor=blue]{hyperref}
\usepackage{latexsym, epsfig, amssymb, amsmath, amsthm, graphicx, mathrsfs, booktabs}
\usepackage{rotating, tabularx, setspace,paralist}
\usepackage{multirow,multicol}
\usepackage{threeparttable}
\def \bfm#1{\mbox{\boldmath$#1$}}

  \def \0 {{\bfm 0}}

\def \tr {\mbox{tr}}

 \def \cd {{\mathcal D}}

\newcommand{\ea}{\end{array}}
\newcommand{\bt}{\begin{tabular}}
\newcommand{\et}{\end{tabular}}
\newcommand{\btb}{\begin{table}}
\newcommand{\etb}{\end{table}}
\newcommand{\ec}{\end{center}}
\newcommand{\bea}{\begin{eqnarray}}
\newcommand{\eea}{\end{eqnarray}}
\newcommand{\Bea}{\begin{eqnarray*}}
\newcommand{\Eea}{\end{eqnarray*}}
\newcommand{\beq}{\begin{equation}}
\newcommand{\eeq}{\end{equation}}

\newcommand{\ignore}[1]{}{}

\newtheorem{theorem}{Theorem}

\newtheorem{corollary}{Corollary}
\newtheorem{proposition}{Proposition}
\theoremstyle{definition}

\newtheorem{example}{Example}

\newtheorem{construction}{Construction}

\begin{document}
\title{\bf Modeling and designs for constrained order-of-addition experiments}

\author{
Jianbin Chen$^1$, Dennis K. J. Lin$^2$, Nicholas Rios$^3$, Xueru Zhang$^4$\thanks{Corresponding author. Email:
zhangxueru2019@gmail.com} \\
$^1$School of Mathematics and Statistics, Beijing Institute of Technology, \\Beijing, BJ 100081, China\\
$^2$Department of Statistics, Purdue University, West Lafayette, \\IN 47907, U.S.A\\
$^3$Department of Statistics, George Mason University, Fairfax, \\VA 22031, U.S.A\\
$^4$School of Mathematics and Physics, University of Science and Technology Beijing, \\Beijing, BJ 100083, China\\
}
\date{}
\maketitle
\begin{quote} \emph{Abstract}: In an order-of-addition (OofA) experiment, the sequence of $m$ different components can significantly affect the response of the experiment. In many OofA experiments, the components are subject to constraints, where certain orders are impossible. For example, in survey design and job scheduling, the components are often arranged into groups, and these groups of components must be placed in a fixed order. If two components are in different groups, their pairwise order is determined by the fixed order of their groups. Design and analysis are needed for these pairwise-group constrained OofA experiments. A new model is proposed to accommodate pairwise-group constraints. This paper also introduces a model for mixed-pairwise constrained OofA experiments, which allows one pair of components within each group to have a pre-determined pairwise order. It is proven that the full design, which uses all feasible orders exactly once, is $D$- and $G$-optimal under the proposed models. Systematic construction methods are used to find optimal fractional designs for pairwise-group and mixed-pairwise constrained OofA experiments. The proposed methods efficiently assess the impact of question order in a survey dataset, where participants answered generalized intelligence questions in a randomly assigned order under mixed-pairwise constraints.


\end{quote}

\maketitle
\section{Introduction}
In an order-of-addition (OofA) experiment, the sequence in which $m$ components are applied impacts the response. The goal of OofA experiments is to determine which sequence of the components is optimal. The optimal order minimizes (or maximizes) the response, though it is also possible to seek an order that brings the response as close as possible to a pre-determined target value. Since there are $m!$ possible permutations of $m$ components, it is too costly to examine all possible orders of addition. For example, when $m = 10$, there are $10!$ (approximately 3.6 million) possible orders. Therefore, a common strategy is to carefully select a design, i.e., a subset of all possible permutations of the components, that is optimal under an appropriately assumed model. Two commonly used models to study OofA experiments are the pairwise ordering (PWO) model of \cite{van1995design} and the component-position model of \cite{yang2021component}. Many recent works have proposed optimal OofA designs with a relatively small number of runs under the aforementioned models and their variants, such as \cite{lin2019order}, \cite{peng2019design}, \cite{voelkel2019design}, \cite{winker2020construction}, \cite{chen2021statistical}, 
\cite{yang2021component}, \cite{zhao2021designs}, \cite{stokes2022position}, \cite{zhao2022optimal}, \cite{chen2022ordering}, and \cite{yang2023ordering}. \cite{chen2021analysis} employed these optimal OofA designs and the PWO model to address heteroscedastic OofA experiments. 
Besides, \cite{lin2023adaptive} applied the quick-sort algorithm to explore the optimal order without any model specification. 

Current research on OofA experiments assumes that all permutations of the 
$m$ components are permissible. However, this is not always the case in practical scenarios. 
For example, consider a single-machine group scheduling problem originally investigated by \cite{yoshida1973study} and \cite{nakamura1978group}, where jobs are arranged into multiple groups and the groups of jobs must be performed in a fixed order. In this case, if two components are in different groups, their pairwise order is determined by the order of their groups. Therefore, these group constraints can be represented as a set of pairwise constraints on the components. Existing methods for OofA experiments cannot directly be applied to pairwise-group constrained OofA experiments. To solve this problem, a new model, called the pairwise-group constrained ordering (PWGCO) model, is proposed. Under this model, the full design, which uses each feasible order under the pairwise-group constraints exactly once, is shown to be optimal with respect to several criteria. To reduce the number of runs, a systematic construction method is presented to construct $\phi$-optimal fractional CODs.





In addition to pairwise-group constraints, there are often constraints on the order of pairs of questions within the same group. For example, a question on a survey can refer to another question within the same group of questions. This happened in a case study using data from George Mason University, which examined the effect of the order of general intelligence questions, as will be detailed in Section \ref{CASE}. This study involved six questions divided into two groups, with the first group always preceding the second. In the second group, Question 5 referred to an image that was shown in Question 4. Therefore, Question 4 always had to be placed before Question 5. This is a mixed-pairwise constrained OofA experiment. We propose a mixed-pairwise constrained ordering (MPWCO) model. It will be shown that the full design is $D$- and $G$-optimal for the proposed model. A systematic construction method is proposed to construct $D$- and $G$-optimal fractional CODs under the MPWCO model, achieving fewer runs than full CODs. 

These experimental designs will allow practitioners to economically and efficiently search for the optimal order of addition, even in cases where there are pairwise-group or mixed-pairwise constraints. This is useful in many fields where it is desirable to find an optimal order of components. In education, \cite{anaya2022understanding} demonstrated that placing very difficult questions at the beginning of an assessment negatively impacts the final PISA scores, indicating the significance of exam question order. The proposed methods can be used to search for the exam question order that maximizes these scores, even if there are constraints on the order of the questions. In survey design, these methods can be used to quickly identify if the order of the questions affects the responses. For example, \citep{lee2009effect} found that asking about self-health before chronic conditions negatively impacted self-rated health scores. These methods are also particularly useful for job scheduling \citep{yoshida1973study} and sequential ordering problems \citep{topcuoglu2002performance}, where a minimum-cost ordering of several jobs needs to be found that satisfies pairwise order constraints.

The remainder of this paper is structured as follows. Section \ref{model} reviews existing models and designs for OofA experiments, and introduces new models for pairwise-group and mixed-pairwise constrained OofA experiments. Section \ref{theory} provides theoretical support for the optimality of full CODs under the proposed models, encompassing all feasible permutations. In Section \ref{COD}, construction methods and illustrative examples are presented for developing optimal fractional CODs based on the proposed models. Section \ref{CASE} applies the proposed model and optimal design to a real case study, analyzing data where survey questions are organized into two groups with a pairwise constraint within the second group. It is shown that the pairwise order of questions has a significant impact on the scores of the survey questions. Extensions for other pairwise constrained OofA experiments are discussed in Section \ref{extensions}. Conclusions are provided in Section \ref{conclusion}. The proofs and the dataset of the real case study are deferred to the Supplementary Material. 

\section{Modeling for pairwise constrained OofA experiments}\label{model}
\subsection{Preliminaries}

Suppose that there are $m$ different components in an OofA experiment, labeled $1, \ldots, m$.  These components can be ordered in $m!$ distinct ways. Each order ${\bfm \pi}=(\pi_1,\ldots, \pi_m)$ is a permutation of $1, \ldots, m$. Let $\Pi_m$ be the set of all possible orders. For any pair of two distinct components $i$ and $j$ in an order ${\bfm \pi}\in \Pi_m$, \cite{van1995design} first proposed ``pseudo factors,'' called pairwise ordering (PWO) factors, which can be written as 
\begin{equation}\nonumber
I_{ij}({\bfm \pi})=\left\{
\begin{aligned}
1& ~~~ {\rm if}~i~{\rm precedes} ~j~in ~{\bfm  \pi},\\
-1& ~~~{\rm if} ~j~ {\rm precedes}~ i ~in ~{\bfm  \pi}.\\
\end{aligned}
\right.
\end{equation}
Under this coding scheme, there are ${m \choose 2}$ different PWO factors.  Without loss of generality, suppose $1\leq i<j\leq m$.  The first-order PWO model is
\begin{equation}\label{e1}
  y=\beta_0+\sum_{i<j}\beta_{ij}I_{ij}(\bfm  \pi)+\epsilon,
\end{equation} where $\bfm\pi \in \Pi_m$, $\beta_{0}$ is the parameter for the constant term, $\beta_{ij}$ is an unknown parameter corresponding to $I_{ij}(\bfm \pi)$,  
and $\epsilon \sim N(0, \sigma^2)$. 

It is challenging to perform an experiment with all possible orders in $\Pi_m$, especially when $m$ is large. For example, when $m$ is 10, there are more than 3.6 million possible orders. This leads to the design problem of selecting a subset of orders to estimate the parameters in Model \eqref{e1}. Various criteria can be applied to measure the optimality of a design by assessing the information it provides. Let $\bfm{x}(\bfm\pi) = (1, I_{12}(\bfm\pi), \dots, I_{(m-1)m}(\bfm\pi))^T$ be the model expansion of a permutation $\bfm \pi \in \Pi_m$. Let the moment matrix be $M(\xi) = \sum_{\bfm\pi \in \Pi_m} \xi(\bfm\pi)\bfm{x}(\bfm\pi)\bfm{x}(\bfm\pi)^T$, where $\xi(\bfm\pi) \geq 0$ and $\sum_{\bfm\pi \in \Pi_m} \xi(\bfm\pi) = 1$. Commonly used criteria include $A$-, $D$- and $G$-criteria, which correspond to $\phi(M)=-\tr\left(M(\xi)^{-1}\right)$, $\log ({\det}(M(\xi)))$, and $\max_{\bfm\pi \in \Pi_m}\{\bfm{x}(\bfm\pi)^TM(\xi)^{-1}\bfm{x}(\bfm\pi)\}$, respectively. These optimality criteria belong to a class of $\phi$-optimality criteria, where $\phi$ must satisfy two conditions: (i) concavity over $M(\xi)$, and (ii) signed permutation invariance, meaning that $\phi\left(R^{\mathrm{T}} M(\xi) R\right)=\phi(M(\xi))$ for every signed permutation matrix $R$. A design measure $\xi$ is $\phi$-optimal if it maximizes $\phi{(M(\xi))}$ among all design measures. Under Model \eqref{e1}, \cite{peng2019design} theoretically proved that the uniform measure over $\Pi_m$ is $\phi$-optimal. This result implies that a full OofA design over $\Pi_m$ is $\phi$-optimal. To reduce the number of runs, \cite{peng2019design} and \cite{chen2020construction} proposed systematic construction methods to construct optimal fractional OofA designs. These designs have significantly smaller run sizes than the full OofA design. 

\subsection{A pairwise-group constrained ordering model}
Suppose that all $m$ components are divided into $n_{G}$ groups in an OofA experiment, denoted by $G_1,\ldots,G_{n_G}$, where $n_G\in \{1,\ldots,m\}$. Each group contains at least one component, and each component occurs only in one group. 
Let $|G_g|$ denote the number of components in group $G_g$, such that $\sum_{g=1}^{n_G}|G_g|=m$. Consider a pairwise-group constraint $G_{g_1} \rightarrow  G_{g_2}$, where all components in $G_{g_1}$ are required to precede all components in $G_{g_2}$ with $1\leq g_1 \not= g_2 \leq n_{G}$. This constraint specifies the relative positions of two distinct groups. An OofA experiment that adheres to the pairwise-group constraints   $G_{1} \rightarrow \cdots \rightarrow G_{n_{G}}$ is called a pairwise-group constrained OofA experiment. Let $\Pi_{(m, G_{1} \rightarrow \cdots \rightarrow G_{n_{G}})}$ be the feasible set of addition orders of components under the constraints $G_{1} \rightarrow \cdots \rightarrow G_{n_{G}}$. This feasible set includes only those orders that comply with the specified pairwise-group constraints. The total number of such feasible orders is $\prod_{g=1}^{n_{G}} |G_{g}|!$. 

When $n_{G}=1$, the OofA experiment follows a traditional format without any constraints on the order. In this case, existing statistical methods can be used to design OofA experiments and search for an optimal order of addition. When $n_{G}=m$, each group only has one component and $\Pi_{(m,  G_{1} \rightarrow \cdots \rightarrow G_{n_{G}})}$ includes only one feasible order, which is inherently the optimal order. However, many practical pairwise-group constrained problems do not fall into these two special cases. For instance, consider a single-machine group scheduling problem as outlined in  \cite{bai2012single}, where $n_{G}=3$ groups of jobs (components) are involved. The structure of such a pairwise-group constrained OofA experiment is displayed in Table \ref{t1}. In this scheduling problem, there are six jobs, denoted as $J_{11},\ldots,J_{32}$. To maintain  consistent notation, $1,\ldots,6$  are used to represent $J_{11},\ldots,J_{32}$, respectively. These jobs are divided into three groups ${G}_1=\{1,2\}$, ${G}_2=\{3,4\}$ and ${G}_3=\{5,6\}$, with pairwise-group constraints $G_{1} \rightarrow G_{2}\rightarrow G_{3}$.

\begin{table}[h!]
\centering
\caption{A single-machine group scheduling problem.}\begin{threeparttable}
\begin{tabular}{ccccccccc}
 \hline \text { Group } & \multicolumn{2}{c}{${G}_1$} && \multicolumn{2}{c}{${G}_2$}&& \multicolumn{2}{c}{${G}_3$}\\
\cline { 2-3 } \cline { 5-6}\cline { 8-9 }
Job & $J_{11}$ & $J_{12}$ && $J_{21}$ & $J_{22}$ && $J_{31}$ & $J_{32}$\\
\hline $p_{i j}$ &25&	30	&&24	&10&&	35&	20\\\hline 
\end{tabular}
{\small \textit{Note: Each job $J_{ij}$ has a given  processing time $p_{ij}$, with  pairwise-group constraints} $G_{1} \rightarrow G_{2}\rightarrow G_{3}$.} 
\end{threeparttable}\label{t1}
\end{table}

Pairwise-group constraints can be equivalently expressed by pairwise constraints of components. For example, $G_{g_1} \rightarrow G_{g_2}$ can be expressed as a set of pairwise constraints of components, i.e., $\mathcal{C}_{G_{g_1} \rightarrow G_{g_2}}=\{i_c \rightarrow  j_c: i_c \in G_{g_1} {\rm and}~j_c \in G_{g_2}\}$ where $i_c \rightarrow  j_c$ denotes that component $i_c$ must precede component $j_c$. Notice that in Model \eqref{e1}, if $i_c \rightarrow j_c$ and $i_c < j_c$, then $I_{i_cj_c}(\bfm \pi) = 1$ for all $\bfm \pi \in \Pi_{(m,  G_{1} \rightarrow \cdots \rightarrow G_{n_{G}})}$. Similarly, if  $i_c \rightarrow j_c$ and $i_c > j_c$, then $I_{j_ci_c}(\bfm \pi) = -1$ for all $\bfm \pi \in \Pi_{(m,  G_{1} \rightarrow \cdots \rightarrow G_{n_{G}})}$. Without loss of generality, we assume that $i_c < j_c$ for all $i_c \in G_{g_1}$ and for all $j_c \in G_{g_2}$, where $1\leq g_1<g_2\leq n_{G}$.




Since $I_{i_cj_c}(\bfm \pi)=1$ for all feasible $\bfm \pi$, the PWO factor $I_{i_cj_c}(\bfm \pi)$ should be removed from the PWO model \eqref{e1}, as it is fully confounded with the constant term. In other words, 
only the PWO factors corresponding to pairwise components within each group need to be considered, while the PWO factors corresponding to pairwise components between different groups should be removed. Based on this, a pairwise-group constrained ordering (PWGCO) model is proposed, which is given by
\bea \label{Deleab_PWO}
y=\beta_0+\mathop{\sum}_{\footnotesize\substack{
    1\leq i<j\leq m \\
    \{i\rightarrow j\} \notin \mathcal{C}_{G_{1} \rightarrow \cdots \rightarrow G_{n_{G} }}}}\beta_{ij}I_{ij}(\bfm  \pi)+\epsilon, \eea
where $\bfm \pi \in \Pi_{(m, G_{1} \rightarrow \cdots \rightarrow G_{n_{G}})}$. The total number of unknown parameters is $1+\sum_{g=1}^{n_{G}}{|G_{g}| \choose 2}$. The following example illustrates Model \eqref{Deleab_PWO}. 



\begin{example}\label{exp1}
For the single-machine group scheduling problem in Table \ref{t1}, the pairwise-constrained set and the feasible set of orders are $\mathcal{C}_{ G_{1} \rightarrow \cdots \rightarrow G_{n_{G}}}=\{1\rightarrow 3,1\rightarrow 4,1\rightarrow 5,1\rightarrow 6,2\rightarrow 3,2\rightarrow 4,2\rightarrow 5,2\rightarrow 6,3\rightarrow 5,3\rightarrow 6,4\rightarrow 5,4\rightarrow 6\}$ and $\Pi_{(6,  G_{1} \rightarrow G_{2}  \rightarrow G_{3})}=\{(1,2,3,4,5,6)$, $(1,2,4,3,5,6)$, $(1,2,3,4,6,5)$, $(1,2,4,3,6,5)$, $(2,1,3,4,5,6)$, $(2,1,4,3,5,6)$, $ (2,1,3,4,6,5)$, $(2,1,4,3,6,5)\}$, respectively. The resulting PWGCO model is
\begin{equation}\nonumber
  y=\beta_0+\beta_{12}I_{12}(\bfm\pi)+\beta_{34}I_{34}(\bfm\pi)+\beta_{56}I_{56}(\bfm\pi)+\epsilon,
\end{equation}
where $\bfm \pi \in \Pi_{(6,  G_{1} \rightarrow G_{2}  \rightarrow G_{3})}$. This model has $1+\sum_{g=1}^3{2 \choose 2}=4$ unknown parameters.
\end{example}
\subsection{A mixed-pairwise constrained ordering model}
When designing a questionnaire, the order of the questions is crucial, as it can influence responses, such as the completion time and the response rate \citep{carter2005iq,lietz2010research,anaya2022understanding}. A key consideration for question ordering is maintaining a logical flow. For example, questions can be divided into different difficulty levels: easy, medium, and difficult. A logical ordering of the questions would be to place easy questions first, followed by medium questions, and then difficult questions. Starting with easy questions helps boost the confidence of the respondents. This ordering of the groups improves the quality of responses and the overall effectiveness of the questionnaire. This question order problem can be viewed as a pairwise-group constrained OofA experiment with pairwise-group constraints $G_1\rightarrow G_2\rightarrow G_3$, where $G_1, G_2$ and $G_3$ represent the easy, medium and difficult questions, respectively. 


In addition to pairwise-group constraints, a logical flow often requires additional pairwise order constraints within groups. For example, a question may require participants to recall a minor detail from another question within the same group. Such an OofA experiment is called a mixed-pairwise constrained OofA experiment, encompassing both inter-group and intra-group constraints. Let  $\mathcal{C}_{G_g}$ be the set of pairwise constraints of components within the group $G_g$. Let $\Pi_{(m,\mathcal{C}_{{G}_{1} \rightarrow \cdots \rightarrow {G}_{n_{{G}}}}\cup\mathcal{C}_{G_{1}} \cup\cdots\cup\mathcal{C}_{G_{n_{G}}})}$ be the feasible set of addition orders of $m$ components under the pairwise-group constraints $G_{1} \rightarrow \cdots \rightarrow G_{n_{G}}$ and pairwise constraints $\mathcal{C}_{G_{1}} \cup\cdots\cup\mathcal{C}_{G_{n_{G}}}$. When $\mathcal{C}_{G_{1}} \cup\cdots\cup\mathcal{C}_{G_{n_{G}}}$ is empty, this notation is simplified to $\Pi_{(m,{G}_{1} \rightarrow \cdots \rightarrow {G}_{n_{{G}}})}$. In this framework, it is assumed that there is at most one pairwise constraint within each group. Therefore, $|\mathcal{C}_{G_g}|$ is either 1 or 0. The total number of such feasible orders is $\prod_{g=1}^{n_{G}} |G_{g}|!/2^{|\mathcal{C}_{G_g}|}$. In Model \eqref{Deleab_PWO}, the PWO factor $I_{i_cj_c}(\bfm \pi)$ should also be removed if there is a pairwise constraint $i_c\rightarrow j_c$ or $j_c\rightarrow i_c$ in $\mathcal{C}_{G_{1}} \cup\cdots\cup\mathcal{C}_{G_{n_{G}}}$. This leads to the following model for mixed-pairwise constrained OofA experiments, \bea \label{Deleab_PWO1}
y=\beta_0+\mathop{\sum}_{\tiny\substack{
    1\leq i<j\leq m \\
    \{i\rightarrow j\} \notin \mathcal{C}_{G_{1} \rightarrow \cdots \rightarrow G_{n_{G} }}\\
    \{i\rightarrow j\} \notin \mathcal{C}_{G_{1}} \cup\cdots\cup\mathcal{C}_{G_{n_{G} }}\\
}}\beta_{ij}I_{ij}(\bfm  \pi)+\epsilon, \eea
 which is called a mixed-pairwise constrained ordering (MPWCO) model. The total number of unknown parameters is $1+\sum_{g=1}^{n_{G}}[{|G_{g}| \choose 2}-|\mathcal{C}_{G_{g}}|]$. 

\begin{example}\label{exp2}
A survey of faculty and students from George Mason University was designed to test general intelligence, memory, and logical reasoning. All survey questions were organized into two groups, $G_1=\{1, 2\}$ and $G_2=\{3,4,5,6\}$, with $G_1 \rightarrow G_2$. Additionally, there was a pairwise constraint $4\rightarrow 5$ within the group $G_2$, where Question 5 asked participants to recall a minor detail from Question 4. The resulting MPWCO model is
\begin{equation}\nonumber
  y=\beta_0+\beta_{12}I_{12}(\bfm\pi)+\beta_{34}I_{34}(\bfm\pi)+\beta_{35}I_{35}(\bfm\pi)+\beta_{36}I_{36}(\bfm\pi)+\beta_{46}I_{46}(\bfm\pi)+\beta_{56}I_{56}(\bfm\pi)+\epsilon,
\end{equation}
where $\bfm \pi \in \Pi_{(6,  \mathcal{C}_{G_{1} \rightarrow G_{2}}\cup\mathcal{C}_{G_{1}} \cup\mathcal{C}_{G_{2}})}$. It has $1+\sum_{g=1}^2[{|G_{g}| \choose 2}-|\mathcal{C}_{G_{g}}|]=7$ unknown parameters. More details about this questionnaire will be provided in Section \ref{CASE}.
\end{example}

\section{Optimality of full pairwise constrained OofA designs}
\label{theory}
\subsection{Full pairwise constrained OofA designs under the PWGCO model}
 In approximate designs, the design weight $\xi(\bfm\pi)$ for an order $\mathbf{\bfm\pi} \in \Pi_{(m, G_{1} \rightarrow \cdots \rightarrow G_{n_{G}})}$ represents the proportion of runs in the design that should be allocated to the order $\bfm \pi$. An exact design requires that $\xi(\bfm \pi)$ is an integer multiple of $1/N$, where $N=\prod_{g=1}^{n_{G}} |G_{g}|!$. The approximate design relaxes this assumption, and allows $\xi(\bfm \pi)$ to be any nonnegative number. Under a design 
 measure $\xi$, the moment matrix is 
 \bea\label{moment}
M(\xi)= \sum_{\bfm \pi}\xi(\bfm \pi)\bfm x(\bfm \pi)\bfm x(\bfm \pi)^T,
\eea
where $\bfm {\pi} \in \Pi_{(m, G_{1} \rightarrow \cdots \rightarrow G_{n_{G}})}$, $\xi(\mathbf{\bfm \pi}) \geq 0$, $\sum_{\mathbf{\bfm \pi}}\xi({\bfm\pi}) = 1$, and $\bfm{x}(\bfm\pi)$ is the expansion of $\bfm\pi$ under Model \eqref{Deleab_PWO}. For example, in Example \ref{exp1}, we have $\bfm{x}({\bfm \pi}) = (1, I_{12}({\bfm \pi}), I_{34}({\bfm \pi}), I_{56}({\bfm \pi}))^T$. The following theorem establishes the optimality of the uniform design measure over $\Pi_{(m, G_{1} \rightarrow \cdots \rightarrow G_{n_{G}})}$ for Model \eqref{Deleab_PWO}.


\begin{theorem}\label{Opti}
Under the PWGCO model in \eqref{Deleab_PWO}, the uniform measure $\xi_0$ on the feasible set $\Pi_{(m,  {G}_{1} \rightarrow \cdots \rightarrow {G}_{n_{{G}}})}$ is $\phi$-optimal, where $\phi(\cdot)$ is concave and signed permutation invariant.
\end{theorem}
Theorem \ref{Opti} indicates that a full pairwise constrained OofA design (COD) on the feasible set
$\Pi_{(m,  {G}_{1} \rightarrow \cdots \rightarrow {G}_{n_{{G}}})}$ is a $\phi$-optimal design, with $N=\prod_{g=1}^{n_{G}} |G_{g}|!$ runs. Moreover, the full COD is $A$-, $D$- and $G$-optimal under Model \eqref{Deleab_PWO}. The values of the $A$-, $D$- and $G$-criteria  with the uniform measure $\xi_0$ are provided in the following theorem. 

\begin{theorem}\label{det1}
Under the PWGCO model \eqref{Deleab_PWO}, the values of the $A$-, $D$- and $G$-criteria  with the uniform measure $\xi_0$ over $\Pi_{m,  G_1 \rightarrow  \cdots \rightarrow  G_{n_G}}$ are  
\begin{equation}\nonumber
\begin{aligned}
&A_{\rm {value}}(\xi_0)=\tr(M(\xi_0)^{-1})=1+\sum_{g=1}^{n_{G}}\frac{3|G_g|(|G_g|-1)^2}{2(|G_g|+1)},\\
&D_{\rm {value}}(\xi_0)=\det(M(\xi_0))=\prod_{g=1}^{n_{G}}\frac{(|G_g|+1)^{|G_g|-1}}{3^{|G_g|(|G_g|-1)/2}} ~{and}\\
&G_{\rm {value}}(\xi_0)=\max_{\bfm\pi}\{\bfm{x}(\bfm\pi)^TM(\xi_0)^{-1}\bfm{x}(\bfm\pi)\}=1+\sum_{g=1}^{n_{G}}{\frac{|G_{g}|(|G_{g}|-1)}{2}},
\end{aligned}
\end{equation}
respectively. 
\end{theorem}

To assess the efficiency of any design measure $\xi$ relative to $\xi_0$, the $A$-, $D$- and $G$-efficiencies are defined as $A_{\rm eff}(\xi)=A_{\rm {value}}(\xi_0)/A_{\rm {value}}(\xi)$, $D_{\rm eff}(\xi)=(D_{\rm {value}}(\xi)/D_{\rm {value}}(\xi_0))^{1/p}$ and $G_{\rm eff}(\xi)=G_{\rm {value}}(\xi_0)/G_{\rm {value}}(\xi)$, respectively. 
The optimal COD has the same values for the $A$-, $D$- and $G$-criteria as a full COD, as established in Theorem \ref{det1}, with $A_{\rm eff}(\xi)=D_{\rm eff}(\xi)=G_{\rm eff}(\xi)=1$. 

A full COD is often impractical, especially when $m$ is large and $n_{G}$ is small. For instance, when $m=10$, $n_G=2$ and $|G_1|=|G_2|=5$, a full COD requires 14,400 runs. It is desirable to systematically construct optimal fractional CODs, which have fewer runs but still have $A$-, $D$-, and $G$-efficiencies equal to 1. The systematic construction of these fractional designs will be discussed in Section \ref{COD}.


\subsection{Full CODs under the MPWCO model}
In the MPWCO model, the moment matrix has the same formulation as in \eqref{moment}, where $\xi$ is a design measure on $\Pi_{(m,\mathcal{C}_{{G}_{1} \rightarrow \cdots \rightarrow {G}_{n_{{G}}}}\cup\mathcal{C}_{G_{1}} \cup\cdots\cup\mathcal{C}_{G_{n_{G}}})}$ and $\bfm{x}(\bfm\pi)$ is the expansion of $\bfm\pi$ under Model \eqref{Deleab_PWO1}. Similar to Theorem \ref{Opti}, the optimality of the uniform design measure over $\Pi_{(m,\mathcal{C}_{{G}_{1} \rightarrow \cdots \rightarrow {G}_{n_{{G}}}}\cup\mathcal{C}_{G_{1}} \cup\cdots\cup\mathcal{C}_{G_{n_{G}}})}$ is shown in the following theorem, when there is at most one pairwise constraint within each group ($|\mathcal{C}_{G_{g}}|\in\{0,1\}$ for $g=1,\ldots,n_{G}$).
\begin{theorem}\label{Opti1}
Under the MPWCO model \eqref{Deleab_PWO1}, the uniform measure over the feasible set 

\noindent$\Pi_{ (m,\mathcal{C}_{{G}_{1} \rightarrow \cdots \rightarrow {G}_{n_{{G}}}}\cup\mathcal{C}_{G_{1}} \cup\cdots\cup\mathcal{C}_{G_{n_{G} }})}$ is $D$- and $G$-optimal.
\end{theorem}
Similar to Theorem \ref{Opti}, Theorem \ref{Opti1} discovers that a full COD over the feasible set

$\Pi_{(m,\mathcal{C}_{{G}_{1} \rightarrow \cdots \rightarrow {G}_{n_{{G}}}}\cup\mathcal{C}_{G_{1}} \cup\cdots\cup\mathcal{C}_{G_{n_{G} }})}$
is a $D$- and $G$-optimal design, with $N=\prod_{g=1}^{n_{G}}|G_{g}|!/2^{|\mathcal{C}_{G_g}|}$ runs. The MPWCO version of Theorem \ref{det1} is provided in the following theorem. 

\begin{theorem}\label{det}
Under the MPWCO model \eqref{Deleab_PWO1}, the values of the $D$- and $G$-criteria with the uniform measure $\xi_0$ on $\Pi_{(m,\mathcal{C}_{{G}_{1} \rightarrow \cdots \rightarrow {G}_{n_{{G}}}}\cup\mathcal{C}_{G_{1}} \cup\cdots\cup\mathcal{C}_{G_{n_{G} })}}$ are  
\begin{equation}\nonumber
\begin{aligned}
&D_{\rm {value}}(\xi_0)=\prod_{g=1}^{n_{G}}\frac{(|G_g|+1)^{|G_g|-1}}{3^{|G_g|(|G_g|-1)/2}} \mbox{and}\\
&G_{\rm {value}}(\xi_0)=1+\sum_{g=1}^{n_{G}}\left[\frac{|G_{g}|(|G_{g}|-1)}{2}-|\mathcal{C}_{G_{g}}|\right],
\end{aligned}
\end{equation}
respectively. 
\end{theorem}

Combing Theorems \ref{det1} and \ref{Opti1}, the proof of Theorem \ref{det} can be straightforwardly obtained, it is thus omitted here. Based on Theorem \ref{det}, we are able to quickly calculate the $D$- and $G$-efficiencies to assess any design measure $\xi$ under Model \eqref{Deleab_PWO1}. The optimal COD achieve $D_{\rm eff}(\xi)=G_{\rm eff}(\xi)=1$. If at least two groups have one within-group pairwise constraint, the uniform measure $\xi_0$ is is no longer $A$-optimal. As before, it is impractical to use the full COD under Model \eqref{Deleab_PWO1} due to its large run size. For example, when $m=12$, $n_G=2$,  $|G_1|=|G_2|=6$ and $|\mathcal{C}_{G_1}|=|\mathcal{C}_{G_2}|=1$, a full COD requires 129,600 runs. The systematic construction of optimal fractional CODs will be studied in Section \ref{COD}.

\section{Construction of optimal fractional CODs}\label{COD}
The main idea of systematic construction methods is to find $n$ different orders such that their moment matrix matches that of a full COD under the proposed models. In this section, we propose two systematic construction methods for optimal fractional CODs under the PWGCO and MPWCO models.


\subsection{Construction of $\phi$-optimal fractional CODs under the PWGCO model}\label{COD-se1}
We first introduce the block design and order-of-addition orthogonal array (OofA-OA) because they are basic tools for our construction method \citep{zhao2022optimal}. A block design has symbols $1, \ldots, m \, (m \geq 4)$ and $b$ blocks of size $k \, (<m)$ such that symbol $i$ appears in $r_i$ blocks, and symbols $i$ and $j (\not=i)$ occur together in $\lambda_{i j}$ blocks for $i, j=1, \ldots, m$.  An $n \times m^{\prime}$ array is called an OofA-OA  of strength $t$ for $m$ components if each element of the array is $+1$ or $-1$, and, for every $n \times t$ sub-array of the OofA-OA, the frequencies of all possible $t$-tuples are proportional to those found when all $m!$ runs are considered \citep{voelkel2019design}. A systematic construction method of $\phi$-optimal CODs under Model \eqref{Deleab_PWO} is provided as follows.

\begin{construction}\label{G1} \,
 \begin{enumerate}[Step 1.]
 \item Input the number of components $m$ and $n_{G}$ groups as $G_1,\ldots,G_{n_{G}}$, and set $g=1$.

\item  Let $A$ and $\bar{A}$ denote the OofA-OAs of orders $n_{g} \times k_g (< |G_g|)$ and $n_{g} \times (|G_g| - k_g)$, respectively. Construct a block design with the symbols in $G_g$ and $b_g$ blocks of size $k_g$ denoted as $o_1, \ldots, o_{b_g}$, satisfying $\lambda_{ih} - \lambda_{jh} = (r_i - r_j)/2$ for every distinct $i, j, h$ in $G_g$.

\item For $l = 1, \ldots, b_g$, construct $B_l$ and $C_l$ using $A$ and $\bar{A}$ with the symbols in $o_l$ and $G_g\backslash o_l$, respectively. Then, form a $(2n_{g})\times |G_g|$ array
$$
 D^{(l)}=\left(\begin{array}{cc}
 B_l & C_l\\  
\sim C_l & B_l\\
\end{array}\right),
$$ where $\sim C_l$ denotes the column reversal of $C_l$. Assemble a $(2b_gn_{g}) \times |G_g|$ OofA design $D_g = [D^{(1)}; \ldots; D^{(b_g)}]$.

\item Update $g\leftarrow g+1$ and return to Step 2 until $g=n_G$. Stack all possible $(\bfm{w}_1, \ldots, \bfm{w}_{n_{G}})$ into the design $D$ and output $D$, where $\bfm{w}_g$ is a row of $D_g$. 

 \end{enumerate}
\end{construction}

For each $g$, Steps 2--3 aim to construct small-run OofA-OAs, denoted by $D_g$, with components in $G_g$. Step 2 can be executed smoothly if for each $g$, there exists a block design with the symbols in  $G_g$ and $b_g$ blocks of size $k_g$ that satisfies the condition $\lambda_{ih} - \lambda_{jh} = (r_i - r_j)/2$ for every distinct $i, j, h$ in $G_g$, exists. \cite{chen2020construction} provided two sufficient conditions to satisfy this nonintuitive requirement in Step 2. The first condition is that the block design is a balanced incomplete block design (BIBD). The second condition is that the block design is formed by adding a single symbol, say $\min\{G_g\}$, to every block of $\tilde{d}$, where $\tilde{d}$ is a BIBD with symbols $G_g\backslash\min\{G_g\}$ and $b_g$ blocks. Each symbol appears in $r$ blocks and each pair of distinct symbols appears together in $\lambda$ blocks with $
b_g-3 r+2 \lambda=0.$  Additionally, the condition $
b_g-3 r+2 \lambda=0$ holds true if the BIBD $\tilde{d}$ is derived from a Hadamard matrix of order $4t$, resulting in $b_g=4 t-1, r=2 t-1, \lambda=t-1$. Alternatively, this condition is satisfied if $|G_g| (=2 s)$ is even, and the blocks of $\tilde{d}$ are taken as all $(s-1)$-subsets of $G_g\backslash\min\{G_g\}$. 
\cite{chen2020construction} pointed out that the first sufficient condition entails the second one. Moreover, the OofA-OAs of orders $n_{g} \times k_g$ and $n_{g} \times (|G_g| - k_g)$ can be found using results from \cite{voelkel2019design}, with $n_{g} <|G_g|!$.


\begin{theorem}\label{DDDD}
The COD constructed by Construction \ref{G1} is $\phi$-optimal under Model \eqref{Deleab_PWO}, where $\phi(\cdot)$ is concave and signed permutation invariant.
\end{theorem}

The output COD in Construction \ref{G1} is an $n \times m$ $\phi$-optimal fractional COD under Model \eqref{Deleab_PWO}, with $n=\prod_{g=1}^{n_{G}} 2b_gn_{g}$. Here is an example to illustrate the procedure of Construction \ref{G1}. 

\begin{example}\label{exp3}
Suppose $m=15$ and $n_G=2$ groups with $G_1=\{1,\ldots,8\}$ and  $G_2=\{9,\ldots,15\}$. In Step 2, let $A$ and $\bar A$ be the $n_1=12$-run OofA-OAs of four components of \cite{voelkel2019design}, with components $1,\ldots,4$ and components $5,\ldots,8$, respectively. The block design with the symbols in $G_1$ is constructed using the second condition, where $\tilde{d}$ is derived from a Hadamard matrix of order $8$, with $b_1=7$, $r=3$ and $\lambda=1$. In Step 3, a $168 \times 8$ OofA design $D_1$ is constructed. The value of $g$ is updated to 2, and we return to Step 2. We keep $A$ unchanged with components $9,\ldots,12$ and let $\bar A$ be a $n_2=12$-run full CODs of three components $13,14,15$. 
The block design is the same as $\tilde{d}$ when $g=1$, with symbols in $G_2$. In Step 3, a $168 \times 7$ OofA design $D_2$ is constructed. In Step 4, all possible combinations of rows in $D_1$ and $D_2$ are stacked, resulting in a design $D$, which is a $28,224\times 15$ optimal fractional COD under Model \eqref{Deleab_PWO}.
\end{example}

We construct more fractional CODs using Construction \ref{C1}. When $|G_g|=3,4,5$ and 6, we skip Steps 2 and 3 of constructing fractional CODs and use the existing OofA-OA from \cite{voelkel2019design} as $D_g$. Specifically, the OofA-OAs with 3, 4, 5, and 6 components have 6, 12, 12, and 24 runs, respectively. When $|G_g|>6$, we use the block designs $A$ and $\bar A$ from Example 1 in \cite{chen2020construction} to execute Step 2. Table \ref{t2} presents these CODs and compares them with full CODs in terms of run size and $\phi$-criteria under Model \eqref{Deleab_PWO}. The parameters $n_{{G}}$ and $(|{G}_1|,\ldots,|{G}_{n_{G}}|)$ determine the type of COD. Note that $N$ is the number of runs of the full COD, while $n$  is the number of runs of the proposed COD. Table \ref{t2} also shows the ratio of $n$ to $N$ for each proposed COD. In comparison, the proposed fractional CODs are all $\phi$-optimal under Model \eqref{Deleab_PWO}, and their run sizes are much smaller than those of the full CODs, particularly for large $n_G$ and $m$.
\begin{table}[h!]\centering
\caption{Comparisons between full CODs and fractional CODs by Construction \ref{G1} in terms of run size and $\phi$-optimality criterion under Model \eqref{Deleab_PWO}.}\begin{threeparttable}
\begin{tabular}{l ccccc}\hline
\multicolumn{3}{c}{{Type}}&  {Full COD} & {Fractional COD}& Ratio\\
$m$&$n_{{G}}$& $(|{G}_1|,\ldots,|{G}_{n_{G}}|)$ &$N$ & \textbf{$n$} &$n/N$ \\
\midrule
7& 2 & (3,4)& 144 & 72 & 0.5  \\
8 &2 & (3,5) & 720 & 72 & 0.1  \\
8 &2 &(4,4) & 576 &144 &0.25 \\
9&2 & (4,5) & 2,880 & 144 & 0.05 \\
10 &2 &(4,6)& 17,280 & 288 & 0.017 \\
11& 2 & (4,7) & 120,960 & 2,016 &0.117\\
12&2 & (4,8) & 967,680 & 2,016 &0.002\\
13&2 & (4,9) & 8,709,120 & 5,184 &$<$0.001\\
15&2 & (7,8) & 203,212,800 & 28,224 &$<$0.001\\
13&3 & (3,4,6) & 103,680  &1,728 & 0.017\\
14&3 & (3,4,7) & 725,760& 12,096&0.017 \\
14&3 & (3,5,6) & 518,400 &1,728& 0.003\\
12&3 & (4,4,4) & 13,824 &1,728 &0.125\\
15&3 &(4,5,6) & 2,073,600 &3,456& 0.002\\
17&3 &(4,6,7) & 87,091,200 &48,384&$<$0.001\\
21&3 &(6,7,8) & $>1.463\times10^{11}$ &677,376& $<$0.001\\
15&4 & (3,3,4,5) & 103,680 &5,184&0.05\\
18&4 &(3,4,5,6) & 12,441,600 &20,736 & 0.002\\
16&4 & (4,4,4,4) & 331,776& 20,736&0.0625\\
22&4 & (4,5,6,7) & $>1.045\times10^{10}$& 580,608&$<$0.001\\
23&4 & (4,5,6,8) & $>8.360\times10^{10}$&580,608&$<$0.001\\
20&5 & (4,4,4,4,4) & 7,962,624&248,832&0.0313\\
23&5 & (4,4,4,5,6) & $>1.194\times10^{9}$&497,664&$<$0.001\\
\hline
\end{tabular}
{\small \textit{Note: All CODs are $\phi$-optimal under Model \eqref{Deleab_PWO}, where $\phi(\cdot)$ is concave and signed permutation invariant.}}
\end{threeparttable}\label{t2}
\end{table}

\subsection{Construction of $D$-optimal fractional CODs under the MPWCO model}
In a mixed pairwise constrained OofA problem, there is at least one group with a pairwise constraint on its components. We begin by studying systematic construction methods for $D$-optimal CODs with one group and one pairwise constraint, i.e.,  $n_G=1$,  $G_1=\{1,\ldots,m\}$ and $\mathcal{C}_{G_1}=\{i_c\rightarrow j_c\}$ where $i_c\not=j_c$ and $i_c, j_c \in \{1,\ldots,m\}$.   The following two construction methods are provided to construct such a $D$-optimal COD for even and odd $m$, respectively.

Under the pairwise constraint $i_c \rightarrow j_c$, Construction \ref{C1} produces an $(m!/(m/2)!)\times m$ fractional COD for even $m$, while  Construction \ref{C2} produces an $(m!/(m-1)/2)!)\times m$ fractional COD for odd $m$. In Step 1 of Construction \ref{C1}, $o_1,\ldots, o_b$
  form a block design by adding a single symbol $i_c$ to each block of a BIBD that includes all $(m/2-1)$-subsets of $\{1,\ldots,m\}\backslash i_c$.  When $m=4$ and 5, the COD using Constructions \ref{C1} and \ref{C2} consists of each order in the feasible set $\Pi_{(m, i_c\rightarrow j_c)}$ exactly once. When $m>5$, the number of runs of the COD using Constructions \ref{C1} and \ref{C2} is smaller than the full COD (that is, $m!/2$). The ratio of $n$ to $N$ decreases as $m$ increases, where $n$ and $N$ are the run sizes of the full COD and the proposed COD, respectively. 


\begin{construction}\label{C1}
(When $m$ is even)
 \begin{enumerate}[Step 1.]
 \item Input the number of components $m$ and the pairwise constraint $i_c\rightarrow j_c$.
 
 \item Construct $b=\binom{m-1}{m/2-1}$ distinct sets, where each set is comprised of the component $i_c$ and $({m/2-1})$ other components in $\{1.\ldots,m\}\backslash i_c$. These sets are represented by $o_1, \ldots, o_{b}$.

 
   \item For each $l=1,\ldots,b$, let $B_l$ be an $(m/2)!\times (m/2)$ array with rows formed by all permutations of $o_l$ if $j_c \notin o_l$. Otherwise, let $B_l^c$ be an $((m/2)!/2)\times (m/2)$ array with rows formed by all permutations of $o_l$ under the pairwise constraint $i_c \rightarrow j_c$ and construct $B_l=[B_l^c;B_l^c]$. 
   
  \item For each $l=1,\ldots,b$, let $C_l$ be an $(m/2)!\times (m/2)$ array with rows formed by all permutations of $\{1,\ldots,m\}\backslash o_l$ and $\sim C_\ell$ be the column reversal of $C_l$.
  \item For each $l=1,\ldots,b$, define 
  \begin{eqnarray}\nonumber    
D_l=\left\{
\begin{aligned}
&\left(\begin{array}{cc}  
    B_l & C_l\\  
\sim C_l & B_l\\
  \end{array}\right)& {\rm if~~} j_c\in o_l,\\
&\left(\begin{array}{cc}  
   B_l & C_l\\  
   B_l & \sim C_l\\
  \end{array}\right)& {\rm if~~} j_c\notin o_l.\\
\end{aligned}
\right.
\end{eqnarray}
\item  Let  $D=[D_1;\ldots;D_{b}]$ and output $D$. 
 \end{enumerate}
\end{construction}

\begin{construction}\label{C2}
(When $m$ is odd)
 \begin{enumerate}[Step 1.]
  \item Input the number of components $m$ and the pairwise constraint $i_c\rightarrow j_c$.
 \item Randomly select an $i\in \{1,\ldots, m\}\backslash\{i_c,j_c\}$.
 \item Let $D$ be  an $((m-1)!/(m-1)/2)!)\times (m-1)$ COD under the pairw constraint $1 \rightarrow  2$ using Construction \ref{C1}. Change the component labels $1,\ldots,m-1$ of $D$ to $i_c,j_c,\pi_3,\ldots,\pi_{m-1}$  where $\pi_3,\ldots,\pi_{m-1}$ is a permutation of $\{1,\ldots,m\}\backslash\{i_c,j_c, i\}$. 
\item For each $k=1,\ldots,m-1$, add a column of the component $i$ before the $k^{\rm th}$ column of $D$ and obtain $D_k$. Then, obtain $D_m$ by adding a column of the component $i$ after the last column of $D$. 

\item Update $D\leftarrow [D_1;\ldots;D_{m}]$ and output $D$. 
 
 \end{enumerate}
 \end{construction}

\begin{theorem}\label{D2-D3}
The CODs constructed by Constructions \ref{C1} and \ref{C2} are $\phi$-optimal under Model \eqref{Deleab_PWO1}, where $\phi(\cdot)$ is concave and signed permutation invariant.
\end{theorem}
The CODs in Constructions \ref{C1} and \ref{C2} have $A_{\rm eff}=D_{\rm eff}=G_{\rm eff}=1$ under Model \eqref{Deleab_PWO1}. Here is an example to illustrate Constructions \ref{C1} and \ref{C2}. 

  \begin{example}
Assume $m=6$ and $(i_c, j_c)=(1, 3)$. 
In Step 2 of Construction \ref{C1}, construct $o_1=\{1,2,3\}, o_2=\{1,2,4\}, o_3=\{1,2,5\}, o_4=\{1,2,6\}, o_5=\{1,3,4\}, o_6=\{1,3,	5\}, o_7=\{1,3,6\}, o_8=\{1,4,5\}, o_9=\{1,4,6\}$, and $o_{10}=\{1,5,6\}$. Following Steps 3--5, under the pairwise constraint $1\rightarrow 3$, the $120\times 6$ optimal COD under Model \eqref{Deleab_PWO1} is represented in Table \ref{table3}. 

Now suppose $m=7$ and $(i_c, j_c) = (1,3)$. In Step 2 of Construction \ref{C2}, select $i=7$ from $\{2, 4, 5, 6, 7\}$. In Steps 3 and 4, a $840\times 7$ optimal COD under Model \eqref{Deleab_PWO1} is constructed by adding a column of $7$ behind the $k^{\rm th}$ column of the COD in Table \ref{table3} for $k = 1,\ldots,6$ to form $D_1,\dots,D_6$, and then adding a column of $7$ after the last column of the COD in Table \ref{table3} to obtain $D_7$.
\end{example}

Based on Constructions \ref{G1}, \ref{C1} and \ref{C2}, we provide the following construction method for the mixed-pairwise constrained OofA experiment. 

\begin{construction}\label{G2}  \,
 \begin{enumerate}[Step 1.]
 \item Input the number of components $m$ and $n_{G}$ groups as $G_1,\ldots,G_{n_{G}}$, the inter-pairwise-group constraints as $\mathcal{C}_{G_{1}}, \ldots,\mathcal{C}_{G_{n_{G}}}$ and set $g=1$.

\item If $\mathcal{C}_{G_{g}}=\emptyset$, use the Steps 2 and 3 of Construction \ref{G1} to construct $D_g$. If $|\mathcal{C}_{G_{g}}|=1$, replace $m$ and labels $1,\ldots,m$ by $|G_g|$ and $G_g$, respectively. Under the pairwise constraint $\mathcal{C}_{G_{g}}$, construct $D_g$ using Construction \ref{C1} if $|G_g|$ is even, and using Construction \ref{C2} if $|G_g|$ is odd. 
\item Update $g\leftarrow g+1$ and return to Step 2 until $g=n_G$. Stack all possible $(\bfm{w}_1, \ldots, \bfm{w}_{n_{G}})$ into the design $D$ and output $D$, where $\bfm{w}_g$ is a row of $D_g$. 

 \end{enumerate}
\end{construction}

\begin{corollary}\label{coro1}
The COD constructed by Construction \ref{G2} is $D$-optimal and $G$-optimal under Model \eqref{Deleab_PWO1}.
\end{corollary}

Combing Theorems \ref{Opti1} and \ref{D2-D3}, the proof of Corollary \ref{coro1} can be straightforwardly obtained,
it is thus omitted here. 
The COD in Construction \ref{G2} has $D_{\rm eff}=G_{\rm eff}=1$ under Model \eqref{Deleab_PWO1}. Construction \ref{G2} integrates Constructions \ref{G1}, \ref{C1} and \ref{C2}, which can handle more types of CODs. To illustrate its efficiency, we construct more fractional CODs using Construction \ref{G2}, listed in Table \ref{tt1}. The setups of OofA-OAs and the block design
used in Construction \ref{G2} are the same as we discussed in Section \ref{COD-se1}. In Table \ref{tt1}, the parameters $n_{{G}}$, $(|{G}_1|,\ldots,|{G}_{n_{G}}|)$ and $(|\mathcal{C}_{{G}_1|},\ldots,|\mathcal{C}_{{G}_{n_{G}}}|)$ determine the type of COD. The third column of Table \ref{tt1} indicates how many components were allocated to each group. The fourth column of Table \ref{tt1} indicates which groups had an additional pairwise constraint. The value of 1 represents the existence of an additional pairwise constraint, while 0 represents its absence. The fractional CODs for the components $m$ were found in the $n_G$ groups, where $m$ was as large as 38, and $n_G = 1,2,3,4,5$.

Table \ref{tt1} shows the number of runs $N$ in the full COD, and compares it to the number of runs $n$ in the proposed fractional CODs, which are all $D$- and $G$-optimal under Model \eqref{Deleab_PWO1}.  The proposed CODs have much smaller run sizes than the full COD, particularly for large $n_G$ and $m$. For example, when there were $m \geq 24$ components, the proposed fractional CODs always use less than 1\% of all available runs. 
Construction \ref{G2} can accommodate any $(|{G}_1|,\ldots,|{G}_{n_{G}}|)$ and $(|\mathcal{C}_{{G}_1|},\ldots,|\mathcal{C}_{{G}_{n_{G}}}|)$. 
Due to space limitations, the proposed fractional CODs outside of Table 3 are omitted here. However, these unshown fractional CODs often have smaller ratios of $n/N$ compared to those of Table \ref{tt1}.

\begin{table}[h!]
\caption{Comparisons between full CODs and fractional CODs in terms
of run size, $D$- and $G$-optimality criteria under Model \eqref{Deleab_PWO1}.}\label{tt1}
\footnotesize
\centering
\tabcolsep=3pt
\begin{threeparttable}
\begin{tabular}{l cccccccc}\hline
\multicolumn{4}{c}{{Type}}&  {Full COD} & {Fractional COD}& Ratio &\multirow{2}{*}{Method}\\
$m$&$n_{{G}}$& $(|{G}_1|,\ldots,|{G}_{n_{G}}|)$ &$(|\mathcal{C}_{{G}_1|},\ldots,|\mathcal{C}_{{G}_{n_{G}}}|)$ &$N$ & \textbf{$n$} &$n/N$ &  \\\hline
{4}  &1 & $(4)$ &$(1)$ & {12}&   12  &  1&  Construction \ref{C1}   \\
{5} &1 & $(5)$ &$(1)$ &{60} & 60 &  1& Construction \ref{C2}     \\
{6}  &1 & $(6)$&$(1)$ &{360} &  120  &  0.333  & Construction \ref{C1}   \\
{7}  &1 & $(7)$&$(1)$ & {$2,520$}  & 840 &  0.333&  Construction \ref{C2}     \\
{8} &1 & $(8)$ &$(1)$ & {$20,160$} &  1,680  &  0.083  &  Construction \ref{C1}  \\
{9} &1 & $(9)$&$(1)$  & {181,440}& 15,120 &  0.083&  Construction \ref{C2}     \\
{10} &1 & $(10)$ &$(1)$ &{1,814,400} &  30,240  &  0.017  &Construction  \ref{C1} \\
{11}  &1 & $(11)$&$(1)$ & {19,958,400} &  332,640  &  0.017  &Construction  \ref{C2} \\
{12}  &1 & $(12)$&$(1)$ &{239,500,800} & 665,280& 0.003  & Construction  \ref{C1} \\ 
{13} &1 & $(13)$ &$(1)$ &3,113,510,400 &   8,648,640  &  0.003  &  Construction \ref{C2}  \\
{14} &1 & $(14)$ &$(1)$ & {$>4.358\times 10^{10}$}  &  17,297,280& $<$0.001  &Construction \ref{C1}  \\
7& 2 & (3,4)&$(1,0)$ & 72 & 36 & 0.5 & Construction \ref{G2} \\
8 &2 & (3,5)&$(1,0)$ & 360 & 36 & 0.1 & Construction \ref{G2} \\
9 &2 &(4,5)&$(1,0)$ & 1440 &144 &0.1 & Construction \ref{G2}\\
12&2 & (4,8) &$(1,0)$ & 483,840 & 2,016 &0.004& Construction \ref{G2}\\
12&2 & (4,8) &$(1,1)$ & 241,920 & 20,160 &0.083& Construction \ref{G2}\\
15&2 & (7,8) &$(1,0)$& 101,606,400 & 141,120 &0.001& Construction \ref{G2}\\
15&2 & (7,8) &$(1,1)$& 50,803,200 & 28,224 &0.028& Construction \ref{G2}\\
15 &3&(4,5,6) &$(1,0,0)$&1,036,800 & 3456 & 0.003& Construction \ref{G2}\\
15 &3&(4,5,6) &$(1,1,0)$&518,400 & 17,280 & 0.033& Construction \ref{G2}\\
16 &3&(4,5,7) &$(1,0,0)$& 7,257,600 & 24,192 & 0.003& Construction \ref{G2}\\
16 &3&(4,5,7) &$(1,1,0)$&3,628,800 & 120,960 & 0.033& Construction \ref{G2}\\
19&3 &(4,7,8) &$(1,0,0)$& 2,438,553,600 &338,688& $<$0.001& Construction \ref{G2}\\
19&3 &(4,7,8) &$(1,1,0)$& 1,219,276,800 &1,693,440& 0.001& Construction \ref{G2}\\
19&3 &(4,7,8) &$(1,1,1)$& 609,638,400 &16,934,400& 0.028& Construction \ref{G2}\\
24&3 &(7,8,9) &$(1,0,0)$& $>3.687\times10^{13}$ &60,963,840& $<$0.001& Construction \ref{G2}\\
24&3 &(7,8,9) &$(1,1,0)$& $>1.843\times10^{13}$ &609,638,400& $<$0.001& Construction \ref{G2}\\
28&4 & (4,7,8,9) &$(1,0,0,0)$& $>8.849\times10^{14}$& 146,313,216&$<$0.001& Construction \ref{G2}\\
28&4 & (4,7,8,9) &$(1,1,0,0)$& $>4.424\times10^{14}$& 731,566,080&$<$0.001& Construction \ref{G2}\\
28&4& (4,7,8,9) &$(1,1,1,0)$& $>2.212\times10^{14}$& 7.315$\times10^{9}$&$<$0.001& Construction \ref{G2}\\
30&4 & (6,7,8,9) &$(1,0,0,0)$& $>2.654\times10^{16}$& 1,463,132,160&$<$0.001& Construction \ref{G2}\\
30&4 & (6,7,8,9) &$(1,1,0,0)$& $>1.327\times10^{16}$& 7,315,660,800&$<$0.001& Construction \ref{G2}\\
30&4& (6,7,8,9) &$(1,1,1,0)$& $> 6.636\times10^{15}$& $7.315\times10^{10}$&$<$0.001& Construction \ref{G2}\\
38&5 & (4,7,8,9,10) &$(1,0,0,0,0)$& $>3.211\times10^{21}$& $6.320\times10^{10}$&$<$0.001& Construction \ref{G2}\\
38&5 & (4,7,8,9,10) &$(1,1,0,0,0)$& $>1.605\times10^{21}$& $3.160\times10^{11}$&$<$0.001& Construction \ref{G2}\\
\hline
\end{tabular}
{\small \textit{Note: All CODs are $D$- and $G$-optimal under Model \eqref{Deleab_PWO1}.}}
\end{threeparttable}
\end{table}

\section{Case study}\label{CASE}

One case study focused on how the order in which the exam questions were asked impacted the final score of the participants. In the study, participants from George Mason University were asked a series of $m = 6$ questions in a randomized order (with mixed constraints) to test their general intelligence, memory, and logical reasoning. Based on how the data were collected, the questions were organized into two groups ($G_1, G_2$), $G_1 \rightarrow G_2$.

The first group of questions $G_1$ consisted of two questions asked from \cite{carter2005iq}. In the second group $G_2$, there were four questions, three of which were directly adapted from \cite{anaya2022understanding}. The additional question (Question 5) was a memory question that asked participants to recall a minor detail from an image in Question 4. Therefore, there was an additional pairwise constraint $4 \rightarrow 5$. Accounting for all constraints, there were 24 feasible orders in which the questions were asked. Participants took an online survey on Qualtrics, and were randomly assigned one of the 24 orders with uniform probability. The response variable was their score; i.e., the number of correctly answered questions. The questions are shown in the supplementary materials. 

In general, the data set had 66 participants. The time taken to answer the questions was also recorded. The median time it took to answer all 6 questions was 8.46 minutes. Three of the 66 participants took over 2 hours to complete the questions, and these were considered outliers and excluded from the sample.  To illustrate the efficacy of the uniform design, two random samples of size 24 were taken, where one response was randomly selected for each of the 24 possible question orders. The full dataset is provided in the supplementary materials.

\begin{table}[!ht]
\caption{Summary of Model (\ref{Deleab_PWO1}) for two random samples. Bold values indicate significance at the $\alpha = 0.10$ level.}
\begin{subtable}{.5\linewidth}
\caption{Random Sample 1}
\label{tab:randomsample123}
\begin{tabular}{ccccc}
  \hline
 & Estimate & SE & $t$ & $p$ \\ 
  \hline
$\beta_0$ & 3.400 & 0.401 & 8.481 & \textbf{0.000} \\ 
  $\beta_{12}$ & 0.583 & 0.299 & 1.952 & \textbf{0.068} \\ 
  $\beta_{34}$ & 0.250 & 0.401 & 0.624 & 0.541 \\ 
  $\beta_{35}$ & 0.150 & 0.401 & 0.374 & 0.713 \\ 
  $\beta_{36}$ & 0.050 & 0.401 & 0.125 & 0.902 \\ 
  $\beta_{46}$ & -0.700 & 0.401 & -1.746 & \textbf{0.099} \\ 
  $\beta_{56}$ & 0.650 & 0.401 & 1.621 & 0.123 \\ 
   \hline
\end{tabular}
\end{subtable}
\begin{subtable}{0.5\linewidth}
\caption{Random Sample 2}
\label{tab:randomsample1234}
\begin{tabular}{ccccc}
  \hline
 & Estimate & SE & $t$ & $p$ \\ 
  \hline
$\beta_0$ & 3.725 & 0.380 & 9.806 & \textbf{0.000} \\ 
  $\beta_{12}$ & 0.208 & 0.283 & 0.736 & 0.472 \\ 
  $\beta_{34}$ & -0.075 & 0.380 & -0.197 & 0.846 \\ 
  $\beta_{35}$ & 0.275 & 0.380 & 0.724 & 0.479 \\ 
 $\beta_{36}$ & -0.025 & 0.380 & -0.066 & 0.948 \\ 
  $\beta_{46}$ & -1.075 & 0.380 & -2.830 & \textbf{0.012} \\ 
  $\beta_{56}$ & 0.575 & 0.380 & 1.514 & 0.148 \\ 
   \hline
\end{tabular}
\end{subtable}
\end{table}


Tables \ref{tab:randomsample123} and \ref{tab:randomsample1234} show that in both random samples, the pairwise order of Questions 4 and 6 was significant at the level $\alpha = 0.10$. In Table \ref{tab:randomsample123}, the pairwise order of the two questions in $G_1$ (1 and 2) was also significant at the $\alpha = 0.10$ level. Fitting the same model for the full dataset reveals that both of these effects are significant, so this result makes sense. In Tables \ref{tab:randomsample123} and \ref{tab:randomsample1234}, since $\hat{\beta}_{46} < 0$, placing Question 4 before Question 6 has a negative impact on the score of the participants. Since $\hat{\beta}_{12} > 0$, then placing Question 1 before Question 2 has a positive impact on the expected score of the participants. Therefore, to maximize the expected score of respondents, it is recommended to place Question 1 before 2, and Question 6 before Question 4. Both random samples had fairly normal residuals (their Shapiro-Wilk normality p-values are $0.919$ and $0.882$, respectively). 

\section{Extensions for other pairwise constrained OofA experiments}
\label{extensions}
\subsection{Other pairwise constrained OofA experiments}
In practical applications, there may be multiple pairwise constraints within a group in mixed-pairwise constrained OofA experiments. For example, the sequential ordering problem (SOP) involves finding the shortest Hamiltonian path in a graph while adhering to given precedence constraints regarding the order in which the nodes are visited \citep{ascheuer1993cutting}. These precedence constraints can be represented as a class of pairwise constraints among nodes.

Figure 1 of \cite{LETCHFORD201674} illustrates a four-node SOP under two pairwise constraints, $A\rightarrow B$ and $C\rightarrow D$, with nodes labeled $A, B, C$ and $D$. The left half of Table \ref{questions} shows all six possible orders for this case, with $A, B, C$ and $D$ replaced by $1,2,3$ and 4. If we change the constraint $3\rightarrow 4$ to $1\rightarrow 3$, all possible orders that comply with the new pairwise constraints $\mathcal{C}_{G_1}=\{1\rightarrow 2, 1\rightarrow 3\}$ are listed in the right part of Table \ref{questions}, resulting in eight different orders. Despite having the same number of pairwise constraints, the two pairwise constrained OofA experiments exhibit different total numbers of possible orders.

\begin{table}[h]
\centering \caption{Two cases of four-node SOP problems under two pairwise constraints.} \label{t21}
\begin{tabular}{c|cccccc|cccccc}\hline
\multirow{3}{*}{Run}& \multicolumn{6}{c|}{$\Pi_{(4,\mathcal{C}_{G_1})}$}  &  \multicolumn{6}{c}{$\Pi_{(4,\mathcal{C}_{G_1})}$}    \\
& \multicolumn{6}{c|}{$\mathcal{C}_{G_1}=\{1\rightarrow 2, 3\rightarrow 4\}$}  &  \multicolumn{6}{c}{$\mathcal{C}_{G_1}=\{1\rightarrow 2, 1\rightarrow 3\}$}    \\\cline{2-13}
&& $\pi_1$ &  $\pi_2$&$\pi_3$ &  $\pi_4$& & &$\pi_1$ & $\pi_2$& $\pi_3$   & $\pi_4$& \\  \hline         
1  &&  1 &   2  &  3  &  4 &&& 1  &   2  &   3 &    4&\\
2& & 1    & 3    & 2   &  4 && &1  &  2  &  4 &   3 &\\
3 & &1    & 3  &   4  &   2&& & 1   & 3   & 2  &  4 &\\
4   && 3     &1   &  2 &    4& &&  1  &  3 &   4  &  2&\\
5 & &3  &   1  &   4 &    2 & &  &1 &   4 &   2 &   3 &\\
6 && 3    & 4   &  1  &   2  & &&1  &  4 &   3  &  2&\\
7&&&&& & && 4  &  1  &  2  &  3&\\
8 &&&&& & & &4 &   1 &   3 &   2&\\\hline
\end{tabular}\label{questions}
\end{table}


\subsection{Modeling and optimal CODs for other pairwise constrained OofA experiments}

An extension to more general constrained OofA experiments can be made by allowing for more than one pairwise constraint within each group, as long as there is some order that satisfies all the constraints. For example, suppose that group $G_1=\{1,2,3,4\}$ has three pairwise constraints $\{1 \rightarrow 2, 2 \rightarrow 3,  2 \rightarrow 4\}$. If we change $2 \rightarrow 4$ to $3 \rightarrow 2$, a conflict arises because no order can satisfy  $\{1 \rightarrow 2, 2 \rightarrow 3,  3 \rightarrow 2\}$. To study such a pairwise constrained OofA experiment, Model \eqref{Deleab_PWO1} remains applicable if $\mathcal{C}_{G_{g}}$ allows any pairwise constraint of components within the group $G_{g}$ for $g=1,\ldots,n_{G}$. In this case, the PWO factor $I_{i_cj_c}(\bfm \pi)$ should be removed if there is a pairwise constraint $i_c\rightarrow j_c$ or $j_c\rightarrow i_c$ in $\mathcal{C}_{G_{1}} \cup\cdots\cup\mathcal{C}_{G_{n_{G}}}$. Here is an example.

\begin{example}\label{general}
Consider the four-node SOP in the left part of Table \ref{tt1}, which has $m=4$, $n_G=1$, and $\mathcal{C}_{G_1}=\{1\rightarrow 2, 3\rightarrow 4\}$. The relationship between the order $\bfm \pi$ of the nodes and the response can be modeled as follows: \bea \label{exam5}y=\beta_0+\beta_{13}I_{13}(\bfm \pi)+\beta_{14}I_{14}(\bfm \pi)+\beta_{23}I_{23}(\bfm \pi)+\beta_{24}I_{24}(\bfm \pi)+\epsilon.\eea This model contains a total of five unknown parameters.
\end{example}
For a pairwise constrained OofA experiment, the uniform design measure is not always optimal. For example, the $D$-optimal design measure under Model \eqref{general} corresponding to runs 1--6 in the left part of Table \ref{tt1} is $(0.20, 0.15, 0.15, 0.15, 0.15, 0.20
)$, respectively. In terms of a $D$-optimal exact design, the numbers of replicates for runs 1--6 is $(4, 3, 3, 3, 3, 4)$, if the experimental budget allows for twenty trials.

\section{Concluding remarks}\label{conclusion}
This paper systematically studies the modeling and design of pairwise constrained OofA experiments. We introduce a new model, called the pairwise-group constrained ordering model, to analyze the pairwise-group constrained OofA experiments. Their components are often arranged
into groups, and these groups of components must be placed in a fixed order. The full COD, which includes all possible orders that satisfy the  pairwise-group constraints, is shown to be $\phi$-optimal for any optimality criterion, where $\phi(\cdot)$ is concave and signed permutation invariant. A systematic construction method is presented to construct $\phi$-optimal fractional CODs for any number of components and groups. A mixed-pairwise constrained ordering model is proposed to deal with the problem when each group of components may involve a pairwise constraint. Under this model, the uniform measure is shown to be $D$- and $G$-optimal. We also provide a systematic construction method to construct $D$- and $G$-optimal fractional CODs.



The optimal fractional CODs obtained from the proposed construction methods require significantly fewer runs than full CODs. To achieve a greater reduction in the number of runs, we can construct a nearly optimal COD, which slightly lowers the estimation efficiency but has a smaller number of runs. In Construction \ref{G1}, we can replace the OofA-OAs $A$ and $\bar A$ with high-efficiency OofA designs found by \cite{winker2020construction}, who used the threshold accepting algorithm to find many unconstrained OofA designs with high relative $D$-efficiency. For example, \cite{winker2020construction} identified a $7\times 4$ OofA design, which can be used as $A$ and $\bar A$ for the case where $|G_g|=8$, with a $D$-efficiency of 0.890. 
The resulting $D_g$ in Step 3 of Construction \ref{G1} is a $98\times 8$ OofA design with a $D$-efficiency of 0.980. For all cases with $|G_g|=8$ and $\mathcal{C}_{G_g}=\emptyset$ in   Tables \ref{t2} and \ref{tt1}, we use the $98\times 8$ OofA design as  $D_g$ in Step 3 of Construction \ref{G1}, achieving nearly $D$-optimal fractional CODs under Models
 \eqref{Deleab_PWO} and \eqref{Deleab_PWO1}. The number of runs and $D$-efficiencies of these fractional CODs are presented in Table \ref{tt3}, which demonstrates that these designs have smaller run sizes compared to those in Tables \ref{t2} and \ref{tt1}, while maintaining $D$-efficiencies of at least $98\%$.

\begin{table}[h!]
\caption{Some nearly $D$-optimal fractional CODs under Models
 \eqref{Deleab_PWO} and \eqref{Deleab_PWO1}.}\label{tt3}
\footnotesize
\centering
\tabcolsep=2pt
\begin{threeparttable}
\begin{tabular}{l ccccccccc}\hline
\multicolumn{4}{c}{{Type}}&  {Full COD} & {Fractional COD}& Ratio & $D_{\rm eff}$ &\multirow{2}{*}{Method}\\
$m$&$n_{{G}}$& $(|{G}_1|,\ldots,|{G}_{n_{G}}|)$ &$(|\mathcal{C}_{{G}_1}|,\ldots,|\mathcal{C}_{{G}_{n_{G}}}|)$ &$N$ & \textbf{$n$} &$n/N$ & & \\\hline
12&2 & (4,8)& $\emptyset$ & 967,680 & 1,176 &0.001&0.983& Construction \ref{G1}\\
15&2 & (7,8)& $\emptyset$ & 203,212,800 & 16,464 &$<$0.001&0.988& Construction \ref{G1}\\
21&3 &(6,7,8)& $\emptyset$ &$>1.463\times10^{11}$ &395,136& $<$0.001& 0.991&Construction \ref{G1}\\
23&4 & (4,5,6,8) & $\emptyset$& $>8.360\times10^{10}$&338,688&$<$0.001&0.990& Construction \ref{G1}\\
12&2 & (4,8) &$(1,0)$ & 483,840 & 1,176 &0.002&0.983& Construction \ref{G2}\\
15&2 & (7,8) &$(1,0)$& 101,606,400 & 82,320 & $<$0.001&0.988& Construction \ref{G2}\\
19&3 &(4,7,8) &$(1,0,0)$& 2,438,553,600 &197,568& $<$0.001&0.989& Construction \ref{G2}\\
19&3 &(4,7,8) &$(1,1,0)$& 1,219,276,800 &987,840& $<$0.001& 0.989&Construction \ref{G2}\\
24&3 &(7,8,9) &$(1,0,0)$& $>3.687\times10^{13}$ &35,562,240& $<$0.001& 0.993&Construction \ref{G2}\\
28&4 & (4,7,8,9) &$(1,0,0,0)$& $>8.849\times10^{14}$&85,349,376&$<$0.001&0.994 &Construction \ref{G2}\\
28&4 & (4,7,8,9) &$(1,1,0,0)$& $>4.424\times10^{14}$& 426,746,880&$<$0.001&0.994 &Construction \ref{G2}\\
30&4 & (6,7,8,9) &$(1,0,0,0)$& $>2.654\times10^{16}$& 853,493,760&$<$0.001&0.994 &Construction \ref{G2}\\
30&4 & (6,7,8,9) &$(1,1,0,0)$& $>1.327\times10^{16}$& 4,267,468,800&$<$0.001&0.994&Construction \ref{G2}\\
38&5 & (4,7,8,9,10) &$(1,0,0,0,0)$& $>3.687\times10^{21}$& $3.687\times10^{10}$&$<$0.001&0.996& Construction \ref{G2}\\
38&5 & (4,7,8,9,10) &$(1,1,0,0,0)$& $>1.843\times10^{21}$& $1.843\times10^{11}$&$<$0.001&0.996& Construction \ref{G2}\\
\hline
\end{tabular}
{\small \textit{Note: $\emptyset$ represents no pairwise constraints within groups, and $D_{\rm eff}$ represents the $D$-efficiency of the fractional COD.}}
\end{threeparttable}
\end{table}

It is possible to use a similar algorithm to search for a nearly $D$-optimal fractional COD with a small number of runs. These fractional CODs can replace those from Constructions \ref{C1} and \ref{C2}, further reducing the number of runs while maintaining high $D$-efficiency. Additionally, finding an optimal design measure under Model \eqref{Deleab_PWO1} is challenging if $\mathcal{C}_{G_{g}}$ allows any pairwise constraints among the components within the group $G_{g}$. These interesting topics may be worth exploring in future research.

\section*{Acknowledgements}
All authors contributed equally and are listed in alphabetical order.  Chen is supported via the National Natural Science Foundation of China via Grant 12301318. Lin is supported via NSF Grant DMS-18102925.

\bibliographystyle{chicago}
\bibliography{references.bib}

\section*{Appendix}
\emph{Proof of Theorem \ref{Opti}.} 
For any $\bfm \pi=(\pi,\ldots,\pi_m), \bfm a=(a_1,\ldots,a_m) \in \Pi_{(m, G_{1} \rightarrow \cdots \rightarrow G_{n_{G}})}$, we denote a permutation of $\bfm \pi$ in the sequence $\bfm a$ by $\bfm \pi\bfm a=(\pi_{a_1},\ldots,\pi_{a_m})$. For each group $G_g=\{\sum_{i=1}^{g-1} |G_{i}|+1,\ldots,\sum_{i=1}^{g} |G_{i}|\}$,  Lemma A.1 of \cite{peng2019design} indicates that there exists a signed permutation matrix $\tilde R_g(\bfm \pi)$ of order $|G_{g}|\choose 2$ such that $\tilde{\bfm x}_g(\bfm \pi \bfm a)^{\mathrm{T}}=\tilde{\bfm x}_g(\bfm 
 a)^{\mathrm{T}} \tilde R_g(\bfm \pi)$ for every $\bfm \pi, \bfm a \in \Pi_{(m, G_{1} \rightarrow \cdots \rightarrow G_{n_{G}})}$, where
\bea\nonumber\tilde{\bfm x}_g(\bfm a)=(I_{(\sum_{i=1}^{g-1} |G_{i}|+1)(\sum_{i=1}^{g-1} |G_{i}|+2)}(\bfm a),\ldots, I_{(\sum_{i=1}^{g} |G_{i}|-1)(\sum_{i=1}^{g} |G_{i}|)}(\bfm a))^T.\eea
 Taking into account all groups, we denote  ${\bfm x}(\bfm a)^T=(1, \tilde{\bfm x}_1(\bfm a)^T,\ldots,\tilde{\bfm x}_{n_G}(\bfm a)^T)^T$ and $R(\bfm \pi)=\diag\{1, \tilde R_1(\bfm \pi),\ldots,\tilde R_{n_G}(\bfm \pi)\}$. Then, $R(\bfm \pi)$ is a signed permutation matrix of order $\sum_{g=1}^{n_G}{|G_{g}|\choose 2}+1$ and  ${\bfm x}(\cdot)$ includes all variables under Model \eqref{Deleab_PWO}. As a result, we have ${\bfm x}(\bfm \pi \bfm a)^{\mathrm{T}}={\bfm x}(\bfm 
 a)^{\mathrm{T}} R(\bfm \pi)$ for every $\bfm \pi, \bfm a \in \Pi_{(m, G_{1} \rightarrow \cdots \rightarrow G_{n_{G}})}$. 
 
 For a design measure $\xi=\{\xi(\bfm a): \bfm a \in \Pi_{(m, G_{1} \rightarrow \cdots \rightarrow G_{n_{G}})}\}$, let $\bfm \pi \xi$ be the design measure that the weight of $\bfm\pi \bfm a$ is  $\xi(\bfm a)$ for each $\bfm a \in \Pi_{(m, G_{1} \rightarrow \cdots \rightarrow G_{n_{G}})}$. It is shown that $\{\bfm \pi \bfm a: \bfm \pi \in \Pi_{(m, G_{1} \rightarrow \cdots \rightarrow G_{n_{G}})}\}=\{\bfm \pi \bfm a: \bfm a \in \Pi_{(m, G_{1} \rightarrow \cdots \rightarrow G_{n_{G}})}\}=\Pi_{(m, G_{1} \rightarrow \cdots \rightarrow G_{n_{G}})}$ for every fixed $\bfm  a, \bfm  \pi\in \Pi_{(m, G_{1} \rightarrow \cdots \rightarrow G_{n_{G}})}$. Then, we have $$\sum_{\bfm \pi} \bfm x(\bfm \pi \bfm a) \bfm x(\bfm \pi \bfm a)^{T}=\sum_{\bfm a} \bfm x(\bfm a) \bfm x(\bfm a)^{{T}}=M(\xi_0)\sum_{g=1}^{n_G}{|G_{g}|!}$$ where the moment matrix $M(\xi_0)= \diag\left\{1, \tilde M^{(1)}(\xi_0),\ldots,\tilde M^{(n_G)}(\xi_0)\right\}$, $\tilde M^{(g)}(\xi_0)=I_g+V_g/3$, $I_g$ is an identity matrix of order $|G_g|\choose 2$, and $V_g$ is an ${|G_g|\choose 2} \times {|G_g|\choose 2}$ matrix following the formulation in (7) of \cite{peng2019design} if we take $m=|G_g|$. For concave $\phi(\cdot)$, we have \bea \nonumber\frac{1}{\prod_{g=1}^{n_G}{|G_{g}|!}} \sum_{\bfm \pi} \phi(M(\bfm \pi \xi)) &\leq&\phi\left(\frac{1}{\prod_{g=1}^{n_G}{|G_{g}|!}}\sum_{\bfm \pi} M(\bfm\pi \xi)\right) \\\nonumber
 &=&  \phi\left(\frac{1}{\prod_{g=1}^{n_G}{|G_{g}|!}} \sum_{\bfm \pi} \sum_{\bfm a} \xi(\bfm a) \bfm x(\bfm \pi\bfm a) \bfm x(\bfm \pi \bfm a)^T\right)\\\nonumber
&=&  \phi\left(\sum_{\bfm a} \xi(\bfm a) M(\xi_0)\right)\\\nonumber
&=&  \phi\left(M(\xi_0)\right).\\\nonumber
\eea For $M(\bfm \pi \xi)$, it holds that
$$
M(\bfm\pi \xi)=\sum_{\bfm a} \xi(\bfm a) \bfm x(\bfm\pi\bfm a) \bfm x(\bfm \pi \bfm a)^{T}=\sum_{\bfm a} \xi(\bfm a) R(\bfm \pi)^T\bfm x(\bfm a) \bfm x(\bfm a)^{T}R(\bfm\pi)=R(\bfm\pi)^T M(\xi) R(\bfm\pi).
$$ Hence, we have  $\frac{1}{\prod_{g=1}^{n_G}{|G_{g}|!}} \sum_{\bfm \pi} \phi(M(\bfm \pi \xi)) =\phi(M(\xi))\leq\phi(M(\xi_0))$. The result shows that the uniform measure $\xi_0$ in the feasible set $\Pi_{(m,  {G}_{1} \rightarrow \cdots \rightarrow {G}_{n_{{G}}})}$ is $\phi$-optimal under the PWGCO model in \eqref{Deleab_PWO}, where $\phi(\cdot)$ is concave and signed permutation invariant.

\noindent\emph{Proof of Theorem \ref{det1}.} For each group $G_g=\{\sum_{i=1}^{g-1} |G_{i}|+1,\ldots,\sum_{i=1}^{g} |G_{i}|\}$, Theorem 2 of \cite{peng2019design} indicates that $\tilde M^{(g)}(\xi_0)=I_g+V_g/3$ has eigenvalues $1+(|G_g|-2)/3$ and $1/3$, with multiplicities $|G_g|-1$ and $(|G_g|-1)(|G_g|-2)/2$, respectively. Then, the values of $A$-, $D$- and $G$-criteria  with the uniform measure $\xi_0$ over $\Pi_{m,  G_1 \rightarrow  \cdots \rightarrow  G_{n_G}}$ are  
\bea\nonumber A_{\rm {value}}(\xi_0)=\tr\left(M(\xi_0)^{-1}\right)&=&\tr\left(\diag\left\{1,\tilde M^{(1)}(\xi_0),\ldots,\tilde M^{(n_G)}(\xi_0)\right\}^{-1}\right)\\\nonumber
&=&1+\sum_{g=1}^{n_G}\tr\left(\tilde M^{(g)}(\xi_0)^{-1}\right)\\\nonumber
&=&1+\sum_{g=1}^{n_G}\left(1+\frac{|G_g|-2}{3}\right)^{-1}(|G_g|-1)+3\frac{(|G_g|-1)(|G_g|-2)}{2}\\\nonumber
&=&1+\sum_{g=1}^{n_{G}}\frac{3|G_g|(|G_g|-1)^2}{2(|G_g|+1)},\nonumber\eea
\bea\nonumber 
D_{\rm {value}}(\xi_0)=\det(M(\xi_0))&=&\det\left(\diag\left\{1,\tilde M^{(1)}(\xi_0),\ldots,\tilde M^{(n_G)}(\xi_0)\right\}\right)\\\nonumber
&=& \prod_{g=1}^{n_G}\det\left(\tilde M^{(g)}(\xi_0)\right)\\\nonumber
&=&\prod_{g=1}^{n_G}\left(1+\frac{|G_g|-2}{3}\right)^{|G_g|-1}3^{-\frac{(|G_g|-1)(|G_g|-2)}{2}}\\\nonumber
&=&\prod_{g=1}^{n_{G}}\frac{(|G_g|+1)^{|G_g|-1}}{3^{|G_g|(|G_g|-1)/2}} ~{and} \nonumber
\eea
\bea\nonumber 
&&G_{\rm {value}}(\xi_0)\\\nonumber
&=&\max_{\bfm\pi}\left\{\bfm{x}(\bfm\pi)^TM(\xi_0)^{-1}\bfm{x}(\bfm\pi)\right\}\\\nonumber
&=&\max_{\bfm\pi}\left\{\left(1,\tilde{\bfm{x}}_1(\bfm\pi)^T,\ldots,\tilde{\bfm{x}}_{n_G}(\bfm\pi)^T\right)\diag\left\{1,\tilde M^{(1)}(\xi_0),\ldots,\tilde M^{(n_G)}(\xi_0)\right\}^{-1}\right.\\\nonumber
&& \left.\times\left(1,\tilde{\bfm{x}}_1(\bfm\pi)^T,\ldots,\tilde{\bfm{x}}_{n_G}(\bfm\pi)^T\right)^T\right\}\\\nonumber
&=&1+ \sum_{g=1}^{n_G}\tilde{\bfm{x}}_{g}(\bfm\pi)^T\left(\tilde M^{(g)}(\xi_0)\right)^{-1}\tilde{\bfm{x}}_{g}(\bfm\pi)\\\nonumber
&=& 1+\sum_{g=1}^{n_G}{|G_g|\choose 2}\nonumber,
\eea respectively.

\noindent\emph{Proof of Theorem \ref{Opti1}.}
First, we consider the case of $n_G=1$, $G_1=\{1,\ldots,m\}$ and $\mathcal{C}_{G_1}=i_c \rightarrow j_c$. 
In Lemma A.1 of \cite{peng2019design}, we substitute $\bm{x}(\bm{\pi})=(1,I_{12}(\bm{\pi}),\ldots,I_{(m-1)m}(\bm{\pi}))^T$ with $\bm{x}(\bm{\pi})=(I_{12}(\bm{\pi}),\ldots,I_{(m-1)m}(\bm{\pi}))^T$ and $R(\bm{\pi})=\widetilde{R}(\bm{\pi})$, where $\bm{\pi}$ represents a permutation of $1,\ldots,m$. Let $X=(X_{ij})$ be the model matrix where $1\leq i<j\leq m$ and $X_{ij}$ correspond to the parameter $\beta_{ij}$. Consequently, it is evident that Theorem 1 of \cite{peng2019design} remains valid for $\bfm x(\bfm \pi)$. 
Denote the uniform measure and the information matrix by $\omega_0$ and $\tilde{M}(\omega_0)=X^TX/m!$. 
The uniform measure serves as the $\phi$-optimal over $\Pi_m$ follows
\begin{enumerate}
\item  for $1 \leq i<j<k  \leq m$, ${X_{ij}^T X_{ik}}/{N }={X_{ik}^TX_{jk}}/{N}= {1}/{3}$ and $
  {X_{ij}^TX_{jk}}/{N}=-1/3;$
\item  for different $i,j, k,l\in \{1,\ldots,m \}$,
${X_{ij}^TX_{kl}}/{N} =0$ if $i<j$ and $k<l$.
\end{enumerate}


The moment matrix of the uniform design measure $\xi_0$ in $\Pi_{(m, \mathcal{C}_{G_1})}$ under Model \eqref{Deleab_PWO1}, 
is $M(\xi_0)=\sum_{\bfm \pi}\bfm x(\bfm\pi) \bfm x(\bfm \pi)^{T}/N$ where $N=m!/2$ and $\bfm x(\bfm \pi)=(1, I_{12}(\bfm \pi),\ldots,I_{(i_c-1)m}(\bfm \pi), $ $I_{(i_c)(i_c+2)}(\bfm \pi),\ldots, $ $I_{(m-1)m}(\bfm \pi))^T$ if $i_c=j_c-1$, otherwise, $\bfm x (\bfm \pi)=(1, I_{12}(\bfm \pi), \ldots, I_{i_c(j_c-1)}(\bfm \pi), $ $I_{i_c(j_c+1)}(\bfm \pi), \ldots, $$ I_{(m-1)m}(\bfm \pi))^T$. Let $X=(X_0, X_{ij})$ be the model matrix where $1\leq i<j\leq m,~ (i,j)\not=(i_c,j_c)$, $X_{0}$ and $X_{ij}$ correspond to the parameters $\beta_0$ and $\beta_{ij}$, respectively. Then,  $M(\xi_0)=X^TX/N$ satisfies:
\begin{enumerate}
\item for $i<j$,
\begin{equation}   \nonumber   
 \frac{X_{0}^T X_{i j}}{N}= \left\{   
  \begin{array}{ccc}  
    \frac 13, & if ~~i=i_c~~or~~j=j_c,\\  
    -\frac 13, & if ~~i=j_c~~or~~ j=i_c,\\
    0, & otherwise,\\ 
  \end{array}
\right.            
\end{equation} $X_{0}^T X_{0}/N= X_{ij}^TX_{ij}/N=1$;

\item  for $1 \leq i<j<k  \leq m$, \begin{equation}  \nonumber
  \begin{array}{ccc}  \frac{ X_{ij}^T X_{ik}}{N }= \frac{1}{3}, & if ~~ (i,j)\not=(i_c,j_c)~~ or ~~(i,k)\not=(i_c,j_c), \\
  \frac{ X_{ik}^T X_{jk}}{N} =\frac13, & if ~~ (i,k)\not=(i_c,j_c)~~ or ~~(j,k) \not=(i_c,j_c),\\
  \frac{ X_{ij}^T X_{jk}}{N} =-\frac13 & if ~~ (i,j)\not=(i_c,j_c)~~ or ~~(j,k) \not=(i_c,j_c);\\
  \end{array} 
  \end{equation}
\item  for different $i,j, k,l\in \{1,\ldots,m \}$,
${X_{ij}^TX_{kl}}/{N} =0$ if $i<j$, $k<l$ and $(i,j)\not=(i_c,j_c)$ or $(k,l) \not=(i_c,j_c)$.
\end{enumerate}

Denoting the position of $I_{i_cj_c}(\bm\pi)$ in $\bm x(\bm \pi)$ as $k$, we move the $k^{\rm th}$ column of an $m(m-1)/2$-order unit matrix into the first column, and denote the resulting matrix as $P$. Subsequently, it follows that $M(\xi)= P^T\tilde M(\xi)P$ for any measure $\xi$ in $\Pi_{(m, \mathcal{C}_{G_1})} \subset \Pi_{m}$. 
Upon comparing $ M(\xi_0)$ with $\tilde M(\omega_0)$, we obtain $M(\xi_0)=P^T\tilde M(\omega_0) P$. Given that $\phi$ is signed permutation invariant, \cite{peng2019design} demonstrated that $\phi( \tilde M(\omega_0))\geq \phi(\tilde M(\xi))$. Therefore, we conclude that
\bea\nonumber
\phi(M(\xi_0))= \phi(P^T\tilde M(\omega_0)P)= \phi(\tilde M(\omega_0))\geq  \phi(\tilde M(\xi))=\phi(PM(\xi) P^T)=\phi(M(\xi)).
\eea 

 Second, we consider the case $n_G>2$. The moment matrix of the uniform design measure $\xi_0$ in $\Pi_{(m,\mathcal{C}_{{G}_{1} \rightarrow \cdots \rightarrow {G}_{n_{{G}}}}\cup\mathcal{C}_{G_{1}} \cup\cdots\cup\mathcal{C}_{G_{n_{G}}})}$ under Model \eqref{Deleab_PWO1}, 
is $M(\xi_0)=\sum_{\bfm \pi}\bfm x(\bfm\pi) \bfm x(\bfm \pi)^{T}/N$ where $N=\prod_{g=1}^{n_{G}} |G_{g}|!/2^{|\mathcal{C}_{G_g}|}$ and $\bfm x (\bfm \pi)=(1, I_{12}(\bfm \pi),\ldots,I_{(i_c-1)m}(\bfm \pi), I_{(i_c)(i_c+2)}(\bfm \pi),$ $\ldots, I_{(m-1)m}(\bfm x))^T$ if $i_c=j_c-1$, otherwise, $\bfm x(\bfm \pi)=(1, I_{12}(\bfm \pi), \ldots, I_{i_c(j_c-1)}(\bfm \pi), I_{i_c(j_c+1)}(\bfm \pi), \ldots,$ $ I_{(m-1)m}(\bfm \pi))^T$. For each group $G_g=\{\sum_{i=1}^{g-1} |G_{i}|+1,\ldots,\sum_{i=1}^{g} |G_{i}|\}$, let $X^{(g)}=(X_{ij})$ be the model matrix where $i,j\in G_g, i<j, X_{ij}$ corresponds to $\beta_{ij}$. If the pairwise constraint $i_c^{(g)} \rightarrow j_c^{(g)}$ exists, remove $X_{i_c^{(g)}j_c^{(g)}}$ from $X^{(g)}$ where $i_c^{(g)}, j_c^{(g)}\in G_g, i_c^{(g)}<j_c^{(g)}, ~ (i,j)\not=(i_c^{(g)},j_c^{(g)})$. Note that $X_0$ correspond to $\beta_{0}$. Without loss of generality, we assume $(i_c^{(g)}, j_c^{(g)})=(\sum_{i=1}^{g-1} |G_{i}|+1,\sum_{i=1}^{g-1} |G_{i}|+2)$ Then,  we have
\begin{enumerate}
\item for $i<j$ and $i, j \in G_g$,
\begin{equation}   \nonumber   
 \frac{X_{0}^T X_{i j}}{N}= \left\{   
  \begin{array}{ccc}  
    \frac 13, & if ~~i=i_c^{(g)}~~or~~j=j_c^{(g)},\\  
    -\frac 13, & if ~~i=j_c^{(g)}~~or~~ j=i_c^{(g)},\\
    0, & otherwise,\\ 
  \end{array}
\right.            
\end{equation} $X_{0}^T X_{0}/N= X_{ij}^TX_{ij}/N=1$;

\item  for $ i<j<k $ and  $i, j, k, m\in G_g$, \begin{equation}  \nonumber
  \begin{array}{ccc}  \frac{ X_{ij}^T X_{ik}}{N }= \frac{1}{3}, & if ~~ (i,j)\not=(i_c^{(g)},j_c^{(g)})~~ or ~~(i,k)\not=(i_c^{(g)},j_c^{(g)}), \\
  \frac{ X_{ik}^T X_{jk}}{N} =\frac13, & if ~~ (i,k)\not=(i_c^{(g)},j_c^{(g)})~~ or ~~(j,k) \not=(i_c^{(g)},j_c^{(g)}),\\
  \frac{ X_{ij}^T X_{jk}}{N} =-\frac13 & if ~~ (i,j)\not=(i_c^{(g)},j_c^{(g)})~~ or ~~(j,k) \not=(i_c^{(g)},j_c^{(g)});\\
  \end{array} 
  \end{equation}

\item  for  $i,j \in G_{g_1}$, $k,l \in G_{g_2}$, $i<j$, $k<l$ and $g_1\not=g_2$, 
 \begin{equation}  \nonumber
  \begin{array}{ccc}  \frac{X_{ij}^TX_{kl}}{N} == \left\{   
  \begin{array}{ccc}  
   0, & if ~~\{i\not=i_c^{(g_1)}, j\not=j_c^{(g_1)}\}~~ or ~~\{k\not=i_c^{(g_2)}, l\not=j_c^{(g_2)}\},\\ 
    \frac 19, & if ~~(i,k)=(i_c^{(g_1)},i_c^{(g_2)})~~or~~ (j_c^{(g_1)},j_c^{(g_2)}),\\
     -\frac 19, & if ~~(i,k)=(i_c^{(g_1)},j_c^{(g_2)})~~or~~ (j_c^{(g_1)},i_c^{(g_2)}).\\ \end{array}
\right.\\
  \end{array} 
  \end{equation}

\end{enumerate}
 
 Without considering the constant term, let the $M^{-1}_{-1}(\xi_0)$ be the inverse matrix of $M(\xi_0)$. Then, we have $M^{-1}_{-1}(\xi_0)=\diag\{V_{-1}^{(1)}(\xi_0),\ldots,V_{-1}^{(n_G)}(\xi_0)\}$. 
 If $\mathcal{C}_{G_{g}}=\emptyset$, $V_{-1}^{(g)}(\xi_0)=(I_g+V_g/3)^{-1}$ where $I_g+V_g/3$ is shown in the proof of Theorem \ref{det1}. Otherwise, $V_{-1}^{(g)}(\xi_0)=(I_g+V_g/3)^{-1}_{-1}$ where $(I_g+V_g/3)^{-1}_{-1}$ is the matrix obtained by removing the first row and column of $(I_g+V_g/3)^{-1}$. Taking into account the constant term, the inverse matrix of $M(\xi_0)$ is denoted by $M^{-1}(\xi_0)$ as 
$$ {\tiny
\begin{bmatrix}
  a_{0} & \bfm {a^{(1)}}  & \bfm {a^{(2)}}& \dots  &  \bfm a^{(n_G)}\\
  \bfm {b^{(1)}} & (I_1+V_1/3)^{-1}_{-1} &  \bfm 0_{\left({|G_{1}| \choose 2}-|\mathcal{C}_{G_{1}}|\right)
  \times \left({|G_{2}| \choose 2}-|\mathcal{C}_{G_{2}}|\right)} & \dots  & \bfm 0_{\left({|G_{1}| \choose 2}-|\mathcal{C}_{G_{1}}|\right)
  \times \left({|G_{n_G}| \choose 2}-|\mathcal{C}_{G_{n_G}}|\right)}\\
    \bfm {b^{(2)}} &\bfm 0_{\left({|G_{2}| \choose 2}-|\mathcal{C}_{G_{2}}|\right)
  \times \left({|G_{1}| \choose 2}-|\mathcal{C}_{G_{1}}|\right)}&  (I_2+V_2/3)^{-1}_{-1}  & \dots  & \bfm 0_{\left({|G_{2}| \choose 2}-|\mathcal{C}_{G_{2}}|\right)
  \times \left({|G_{n_G}| \choose 2}-|\mathcal{C}_{G_{n_G}}|\right)}\\

   \vdots & \vdots &\vdots & \ddots & \vdots \\
   \bfm b^{(n_G)} &\bfm 0_{\left({|G_{2}| \choose 2}-|\mathcal{C}_{G_{2}}|\right)
  \times \left({|G_{n_G}| \choose 2}-|\mathcal{C}_{G_{n_G}}|\right)}&\bfm 0_{\left({|G_{2}| \choose 2}-|\mathcal{C}_{G_{2}}|\right)
  \times \left({|G_{n_G}| \choose 2}-|\mathcal{C}_{G_{n_G}}|\right)}&\vdots &   (I_{n_G}+V_{n_G}/3)^{-1}_{-1}

\end{bmatrix}}
$$ 
where $a_0=1+ \sum_{g=1}^{n_G}{|\mathcal{C}_{G_{n_G}}|}(c_g-1)$, $c_g$ is an arbitrary diagonal element of $(I_g+V_g/3)^{-1}$, $\bfm a^{(g)}$ is a row vector of elements 0 if $|\mathcal{C}_{G_{g}}|=0$, otherwise the first row of $(I_g+V_g/3)^{-1}$ with the first element removed, and $\bfm b^{(g)}$ is the transpose of $\bfm a^{(g)}$. For any $\bfm \pi \in \Pi_{(m,\mathcal{C}_{{G}_{1} \rightarrow \cdots \rightarrow {G}_{n_{{G}}}}\cup\mathcal{C}_{G_{1}} \cup\cdots\cup\mathcal{C}_{G_{n_{G} })}}$, it holds that $\bfm x(\bfm \pi)^TM^{-1}(\xi_0)\bfm x(\bfm \pi)= 1+\sum_{g=1}^{n_{G}}\left[\frac{|G_{g}|(|G_{g}|-1)}{2}-|\mathcal{C}_{G_{g}}|\right]$. The uniform measure $\xi_0$ is $G$-optimal under Model \eqref{Deleab_PWO1}. According to the equivalence theorem, it is also $D$-optimal under Model \eqref{Deleab_PWO1}.

\noindent\emph{Proof of Theorem \ref{DDDD}.}
According to \cite{chen2020construction}, $D_g$ is a $\phi$-optimal OofA design with $|G_g|$ components in $G_g$, and it has the same moment matrix as $I_g + V_g/3$. In the output $D$, the portion corresponding to group $g$ is repeated $\prod_{i \neq g} |D_i|$ times, and it still maintains the same moment matrix as $I_g + V_g/3$. For $i, j \in G_{g_1}$ and $k, l \in G_{g_2}$ with $g_1 \neq g_2$, we have $X^T_{ij} X_{kl} = 0$ since all possible positions of $i, j, k,$ and $l$ appear equally. For any $i, j \in G_g$, the terms $i \rightarrow j$ and $j \rightarrow i$ appear equally, so $X^T_0 X_{ij} = 0$. The moment matrix of $D$ is the same as that of the uniform measure $\xi_0$ in $\Pi_{(m, \mathcal{C}_{{G}{1} \rightarrow \cdots \rightarrow {G}_{n{G}}})}$, which is an optimal COD.

\noindent\emph{Proof of Theorem \ref{D2-D3}.} 
Let $G_u(i j)$ be the contribution of $D_l$ to $X_0^{\mathrm{T}} X_{i j}$. Then, we have 

\begin{enumerate}
\item $G_u(i_cj) =\left\{   
  \begin{array}{ccc}  
    \frac 23\left(\frac{m}{2}\right)!, & if ~~j \in o_l~~and~~j_c\in o_l,\\  
    0, & if ~~j \in o_l~~and~~j_c\notin o_l,\\
    0, &  if~~j \notin o_l~~and~~j_c\in o_l,\\
     2\left(\frac{m}{2}\right)!, & if ~~j \notin o_l~~and~~j_c\notin o_l;
  \end{array}
\right.            
$
\item $G_u(ij_c) =\left\{   
  \begin{array}{ccc}  
    \frac 23\left(\frac{m}{2}\right)!, & if ~~i \in o_l~~and~~j_c\in o_l,\\  
      2\left(\frac{m}{2}\right)!, & if ~~i \in o_l~~and~~j_c\notin o_l\\
    0, & if ~~i \notin o_l~~and~~j_c\in o_l,\\
    0, &  if~~i \notin o_l~~and~~j_c\notin o_l;\\
  \end{array}
\right.            
$
\item $G_u(j_cj) =\left\{   
  \begin{array}{ccc}  
   - \frac 23\left(\frac{m}{2}\right)!, & if ~~j_c \in o_l~~and~~j\in o_l,\\  
      -2\left(\frac{m}{2}\right)!, & if ~~j_c \notin o_l~~and~~j\in o_l;\\
    0, & if ~~j_c \in o_l~~and~~j\notin o_l,\\
    0, &  if~~j_c \notin o_l~~and~~j\notin o_l;\\
  \end{array}
\right.            
$

\item $G_u(ii_c) =\left\{   
  \begin{array}{ccc}  
   - \frac 23\left(\frac{m}{2}\right)!, & if ~~i \in o_l~~and~~j_c\in o_l,\\  
     0, & if ~~i \in o_l~~and~~j_c\notin o_l;\\
    0, & if ~~i\notin o_l~~and~~j_c\in o_l,\\
     -2\left(\frac{m}{2}\right)!, &  if~~i \notin o_l~~and~~j_c\notin o_l;\\
  \end{array}
\right.            
$

 \item  $G_u(ij) =0$ if $i<j$, $i\not=i_c$ and $j\not=j_c$. 

\end{enumerate}
For case 1, it holds that $$\frac{X_0^TX_{ij}}{n}=\frac{\frac23\left(\frac{m}{2}\right)! \times {m-3\choose m/2-3}+2\left(\frac{m}{2}\right)!\times  {m-3\choose m/2-1}}{2\left(\frac{m}{2}\right)!\binom{m-1}{m/2-1}}=\frac13$$ where $n$ represents the number of runs in the proposed design $D$. Similarly, we can obtain the values of ${X_0^TX_{ij}}/{n}$ for the other four cases. We can see that ${X_0^TX_{ij}}/{n}$ has the same value as the full COD. 

For different $i ,j ,k \in \{1,\ldots,m\}$, let $G_u$ be the row vector with elements $G_u(i j, i k), G_u(i k, j k)$ and $G_u(i j, j k)$, where $G_u(i j, i k)$ is the contribution of $D_l$ to $X_{i j}^{\mathrm{T}} X_{i k}$, and $G_u(i k, j k), G_u(i j, j k)$ are defined similarly. Note that $B_l$ is formed by all permutations of the elements of $o_l$ if $j_c \notin o_l$, otherwise $B_l$ is a two-fold repetition of $\Pi_{(m, i_c\rightarrow j_c)}$. Besides, $C_l$ and $\bar C_l$ are formed by all permutations of the elements of $\{1,\ldots,m\}\backslash o_l$. Then, we have 
 $$G_u =\left\{   
  \begin{array}{ccc}  
    \frac 23\left(\frac{m}{2}\right)!(1,1,-1), & if ~~i,j,k \in o_l ~~or~~ i,j,k \notin o_l,\\  
    2\left(\frac{m}{2}\right)!(1,0,0), & if ~~ i \in o_l, j,k \notin o_l ~~or~~j,k \in o_l, i\notin o_l\\
    2\left(\frac{m}{2}\right)!(0,1,0), &  if~~i,j \in o_l, k\notin o_l ~~or~~k \in o_l, i,j\in o_l,\\
     2\left(\frac{m}{2}\right)!(0,0,-1), & if ~~i,k \in o_l, j\notin o_l ~~or~~j \in o_l, i, k\notin o_l.
  \end{array}
\right.            
$$

Taking into account all the cases, we have $X_{ij}^TX_{ik}/n=X_{ik}^TX_{jk}/n=1/3$ and $X_{ij}^TX_{jk}=-1/3$. This result is the same as that for the full COD. In addition, for different $i,j,k,l \in \{1,\ldots,m\}$, we can similarly obtain $X_{ij}^TX_{kl}/n=0$, which is the same as for the full COD. In conclusion, the COD constructed by Construction \ref{C1} is $\phi$-optimal under Model \eqref{Deleab_PWO1}, where $\phi(\cdot)$ is concave and signed permutation invariant. 

Besides, the optimality of Construction \ref{C2} can be proved similarly and are omitted here.

\begin{table} [h!]
\caption{The optimal COD under Model \eqref{Deleab_PWO1}   with the pairwise constraint $1\rightarrow 3$.}  \label{table3}
\small
\centerline{ \tabcolsep=3truept \bt {ccccccccccccccccccccc}\hline
Run& $\pi_1$   & $\pi_2$  &$\pi_3$ &$\pi_4$ &$\pi_5$  &$\pi_6$  &Run& $\pi_1$   & $\pi_2$  &$\pi_3$ &$\pi_4$ &$\pi_5$  &$\pi_6$ &Run&$\pi_1$    & $\pi_2$  &$\pi_3$ &$\pi_4$ &$\pi_5$  &$\pi_6$\\\hline
 1       &       2       &       1       &       3       &       6       &       5       &       4       &      41       &       1       &       6       &       2       &       3       &       5       &       4       &      81       &       2       &       5       &       4       &       1       &       3       &       6      \\
       2       &       1       &       3       &       2       &       6       &       4       &       5       &      42       &       1       &       2       &       6       &       3       &       4       &       5       &      82       &       5       &       2       &       4       &       6       &       1       &       3      \\
       3       &       1       &       2       &       3       &       5       &       6       &       4       &      43       &       6       &       2       &       1       &       3       &       4       &       5       &      83       &       4       &       5       &       2       &       1       &       6       &       3      \\
       4       &       2       &       1       &       3       &       5       &       4       &       6       &      44       &       6       &       1       &       2       &       4       &       3       &       5       &      84       &       5       &       4       &       2       &       1       &       3       &       6      \\
       5       &       1       &       3       &       2       &       4       &       6       &       5       &      45       &       2       &       6       &       1       &       3       &       5       &       4       &      85       &       5       &       4       &       1       &       6       &       3       &       2      \\
       6       &       1       &       2       &       3       &       4       &       5       &       6       &      46       &       2       &       1       &       6       &       5       &       3       &       4       &      86       &       5       &       1       &       4       &       6       &       2       &       3      \\
       7       &       4       &       5       &       6       &       2       &       1       &       3       &      47       &       1       &       6       &       2       &       4       &       5       &       3       &      87       &       4       &       5       &       1       &       3       &       6       &       2      \\
       8       &       5       &       4       &       6       &       1       &       3       &       2       &      48       &       1       &       2       &       6       &       5       &       4       &       3       &      88       &       4       &       1       &       5       &       3       &       2       &       6      \\
       9       &       4       &       6       &       5       &       1       &       2       &       3       &      49       &       4       &       1       &       3       &       6       &       5       &       2       &      89       &       1       &       5       &       4       &       2       &       6       &       3      \\
      10       &       6       &       4       &       5       &       2       &       1       &       3       &      50       &       1       &       4       &       3       &       6       &       2       &       5       &      90       &       1       &       4       &       5       &       2       &       3       &       6      \\
      11       &       5       &       6       &       4       &       1       &       3       &       2       &      51       &       1       &       3       &       4       &       5       &       6       &       2       &      91       &       5       &       4       &       1       &       2       &       3       &       6      \\
      12       &       6       &       5       &       4       &       1       &       2       &       3       &      52       &       4       &       1       &       3       &       5       &       2       &       6       &      92       &       5       &       1       &       4       &       3       &       2       &       6      \\
      13       &       4       &       2       &       1       &       6       &       5       &       3       &      53       &       1       &       4       &       3       &       2       &       6       &       5       &      93       &       4       &       5       &       1       &       2       &       6       &       3      \\
      14       &       4       &       1       &       2       &       6       &       3       &       5       &      54       &       1       &       3       &       4       &       2       &       5       &       6       &      94       &       4       &       1       &       5       &       6       &       2       &       3      \\
      15       &       2       &       4       &       1       &       5       &       6       &       3       &      55       &       2       &       5       &       6       &       4       &       1       &       3       &      95       &       1       &       5       &       4       &       3       &       6       &       2      \\
      16       &       2       &       1       &       4       &       5       &       3       &       6       &      56       &       5       &       2       &       6       &       1       &       4       &       3       &      96       &       1       &       4       &       5       &       6       &       3       &       2      \\
      17       &       1       &       4       &       2       &       3       &       6       &       5       &      57       &       2       &       6       &       5       &       1       &       3       &       4       &      97       &       6       &       4       &       1       &       5       &       3       &       2      \\
      18       &       1       &       2       &       4       &       3       &       5       &       6       &      58       &       6       &       2       &       5       &       4       &       1       &       3       &      98       &       6       &       1       &       4       &       5       &       2       &       3      \\
      19       &       4       &       2       &       1       &       3       &       5       &       6       &      59       &       5       &       6       &       2       &       1       &       4       &       3       &      99       &       4       &       6       &       1       &       3       &       5       &       2      \\
      20       &       4       &       1       &       2       &       5       &       3       &       6       &      60       &       6       &       5       &       2       &       1       &       3       &       4       &      100      &       4       &       1       &       6       &       3       &       2       &       5      \\
      21       &       2       &       4       &       1       &       3       &       6       &       5       &      61       &       5       &       1       &       3       &       6       &       4       &       2       &      101      &       1       &       6       &       4       &       2       &       5       &       3      \\
      22       &       2       &       1       &       4       &       6       &       3       &       5       &      62       &       1       &       5       &       3       &       6       &       2       &       4       &      102      &       1       &       4       &       6       &       2       &       3       &       5      \\
      23       &       1       &       4       &       2       &       5       &       6       &       3       &      63       &       1       &       3       &       5       &       4       &       6       &       2       &      103      &       6       &       4       &       1       &       2       &       3       &       5      \\
      24       &       1       &       2       &       4       &       6       &       5       &       3       &      64       &       5       &       1       &       3       &       4       &       2       &       6       &      104      &       6       &       1       &       4       &       3       &       2       &       5      \\
      25       &       5       &       2       &       1       &       6       &       4       &       3       &      65       &       1       &       5       &       3       &       2       &       6       &       4       &      105      &       4       &       6       &       1       &       2       &       5       &       3      \\
      26       &       5       &       1       &       2       &       6       &       3       &       4       &      66       &       1       &       3       &       5       &       2       &       4       &       6       &      106      &       4       &       1       &       6       &       5       &       2       &       3      \\
      27       &       2       &       5       &       1       &       4       &       6       &       3       &      67       &       2       &       4       &       6       &       5       &       1       &       3       &      107      &       1       &       6       &       4       &       3       &       5       &       2      \\
      28       &       2       &       1       &       5       &       4       &       3       &       6       &      68       &       4       &       2       &       6       &       1       &       5       &       3       &      108      &       1       &       4       &       6       &       5       &       3       &       2      \\
      29       &       1       &       5       &       2       &       3       &       6       &       4       &      69       &       2       &       6       &       4       &       1       &       3       &       5       &      109      &       6       &       5       &       1       &       4       &       3       &       2      \\
      30       &       1       &       2       &       5       &       3       &       4       &       6       &      70       &       6       &       2       &       4       &       5       &       1       &       3       &      110      &       6       &       1       &       5       &       4       &       2       &       3      \\
      31       &       5       &       2       &       1       &       3       &       4       &       6       &      71       &       4       &       6       &       2       &       1       &       5       &       3       &      111      &       5       &       6       &       1       &       3       &       4       &       2      \\
      32       &       5       &       1       &       2       &       4       &       3       &       6       &      72       &       6       &       4       &       2       &       1       &       3       &       5       &      112      &       5       &       1       &       6       &       3       &       2       &       4      \\
      33       &       2       &       5       &       1       &       3       &       6       &       4       &      73       &       6       &       1       &       3       &       5       &       4       &       2       &      113      &       1       &       6       &       5       &       2       &       4       &       3      \\
      34       &       2       &       1       &       5       &       6       &       3       &       4       &      74       &       1       &       6       &       3       &       5       &       2       &       4       &      114      &       1       &       5       &       6       &       2       &       3       &       4      \\
      35       &       1       &       5       &       2       &       4       &       6       &       3       &      75       &       1       &       3       &       6       &       4       &       5       &       2       &      115      &       6       &       5       &       1       &       2       &       3       &       4      \\
      36       &       1       &       2       &       5       &       6       &       4       &       3       &      76       &       6       &       1       &       3       &       4       &       2       &       5       &      116      &       6       &       1       &       5       &       3       &       2       &       4      \\
      37       &       6       &       2       &       1       &       5       &       4       &       3       &      77       &       1       &       6       &       3       &       2       &       5       &       4       &      117      &       5       &       6       &       1       &       2       &       4       &       3      \\
      38       &       6       &       1       &       2       &       5       &       3       &       4       &      78       &       1       &       3       &       6       &       2       &       4       &       5       &      118      &       5       &       1       &       6       &       4       &       2       &       3      \\
      39       &       2       &       6       &       1       &       4       &       5       &       3       &      79       &       2       &       4       &       5       &       6       &       1       &       3       &      119      &       1       &       6       &       5       &       3       &       4       &       2      \\
      40       &       2       &       1       &       6       &       4       &       3       &       5       &      80       &       4       &       2       &       5       &       1       &       6       &       3       &      120      &       1       &       5       &       6       &       4       &       3       &       2  \\\hline\et}
\end{table}

Survey Questions\\
Group 1: Two general intelligence questions without constraints.\\

Question 1: (Correct Answer: A)\\
\begin{figure}[tph]
\begin{center}
\includegraphics[width=4in]{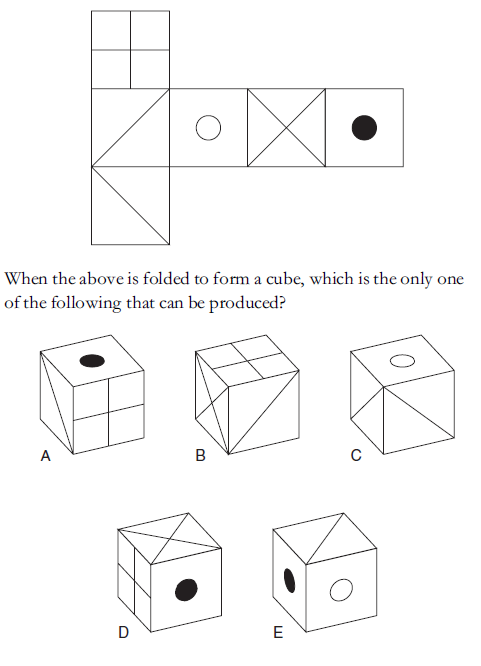}
\end{center}
\end{figure}

Question 2: (Correct Answer: 4)\\
\begin{figure}[tph]
\begin{center}
\includegraphics[width=4in]{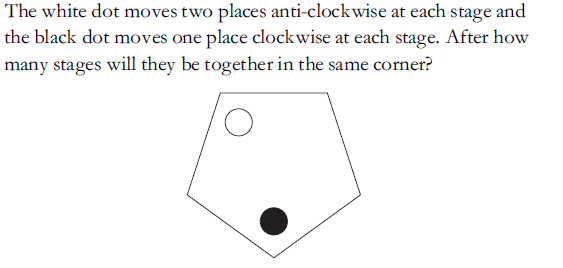}
\end{center}
\end{figure}

Possible Answers:  1 2 3 4 5

Group 2: Four survey questions on math with one pairwise constraint.\\
Question 3: (Easy, Correct Answer: 10)\\
 Luis turns 36 years old today. His age is 9 times that of his cat. The age of his dog is three halves the age of his cat. What is the sum of the ages of the cat and the dog?
 
 Possible Answers: 8 9 10 12 13

Question 4: (Medium, Correct Answer: Three)\\
Ferb has constructed the following figure and, once done, has painted it blue without lifting it off the ground. How many of the cubes have exactly three faces painted blue?
\begin{figure}[tph]
\begin{center}
\includegraphics[width=4in]{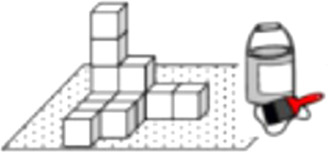}
\end{center}
\end{figure}

Possible Answers: None One Two Three Four

Question 5: (Medium, Correct Answer: (b))\\
What color was the paintbrush handle in the problem with the painted blocks?

Possible Answers: (a) Blue (b) Red (c) Green (d) Yellow (e) The picture was entirely black and white.

Question 6: (Difficult, Correct Answer: 64)\\If we increase the length of all sides of a square by a fixed percentage, and its area increases by 96\%. By what percentage would the area have decreased if instead of increasing the sides, we decreased the sides by the same percentage?\\
Possible Answers: 4\% 64\% 94\% 48\% 36\%

\begin{table} [!t]
\caption{The dataset of the survey questions.}  \label{survey}
\par\vskip .1cm \tiny
\centerline{\tabcolsep=0.5truept \bt {cccccccccccc}\hline
\multirow{2}{*}{Participant}& \multirow{2}{*}{Start Date}&	&\multirow{2}{*}{End Date}& {Duration}&	\multirow{2}{*}{Recorded Date}	&{Distribution}&	\multirow{2}{*}{Score}	&{Group 1}	& Group 2 &	\multirow{2}{*}{Logduration}	& \multirow{2}{*}{Efficiency}\\
& &	& & (seconds)&	&Channel&	&${(\pi_1,\pi_2)}$	& ${(\pi_3,\ldots,\pi_6)}$ &		&\\\hline
1 &	4/11/2024 9:07&&	4/11/2024 9:13&		352	&4/11/2024 9:13	&anonymous&	3	& $(1,2)$& $(6,4,3,5)$&	5.863&	0.511\\
2 &	4/11/2024 10:17	& &4/11/2024 10:22&		266&	4/11/2024 10:22&	anonymous&	6	&	$(1,2)$	& $(3,4,5,6)$ &5.583&1.074\\
3&4/11/2024 10:25&&	4/11/2024 10:27	&	75&	4/11/2024 10:27	&anonymous	&NA&	$(1,2)$	& $(3,4,6,5)$ &4.317&NA\\
4&	4/11/2024 10:27&&	4/11/2024 10:29	&	165	&4/11/2024 10:29	&anonymous&	5	& 	$(1,2)$	&$(3,4,5,6)$&	5.105&0.979\\
5 	& 	4/11/2024 10:26	&&  4/11/2024 10:41& 		896	& 4/11/2024 10:41& 	anonymous	& 3	& $(1,2)$	&$(4,3,5,6)$&6.797&	0.441\\
6& 		4/11/2024 13:07&& 		4/11/2024 13:15	& 		484	& 	4/11/2024 13:15& 		anonymous& 		1	& 	 $(1,2)$	&$(4,3,6,5)$&	6.182&	0.161\\
7&	4/12/2024 8:11&	&4/12/2024 8:19	&	484	&4/12/2024 8:19&	anonymous	&4	&	$(1,2)$	&$(6,4,5,3)$	&6.182&	0.647\\
8&	4/12/2024 8:17&	&4/12/2024 8:22	&	328&	4/12/2024 8:22	&anonymous	&2&	$(2,1)$	&	$(4,	5,	6,	3)$ &	5.793&0.345\\
9 &	4/12/2024 8:11	&&4/12/2024 8:23	&	736&	4/12/2024 8:23	&anonymous	&1	&$(2,1)$	&$(4,	6,	3,	5)$	&	6.601&0.151\\
10&	4/12/2024 8:10	&&4/12/2024 8:25	&	925	&4/12/2024 8:25	&anonymous	&3&	$(2,1)$	&$(6,3,	4,	5)$ &	6.829&	0.439\\
11&	4/12/2024 8:23&&	4/12/2024 8:32	&	569	&4/12/2024 8:32&	anonymous&	5	& 	$(1,2)$&		$(6,4,5,3)$&		6.343&	0.788\\
12&	4/12/2024 8:26&	&4/12/2024 8:32	&	410	&4/12/2024 8:32&	anonymous&	1	& 	$(1,2)$	&$(4,3,6,5)$&6.016&	0.166\\
13&	4/12/2024 8:23&	&4/12/2024 8:36	&	768	&4/12/2024 8:36&	anonymous	&5	&	$(2,1)$	&$(6,3,	4,	5)$	&6.643&	0.752\\
14 &4/12/2024 8:13	&&4/12/2024 8:37	&	1440	&4/12/2024 8:37&	anonymous	&3	&$(1,2)$	&$(3,4,6,5)$	&7.272&	0.412\\
15&	4/12/2024 8:10&	&4/12/2024 8:41	&	1852	&4/12/2024 8:41	&anonymous	&3	&$(1,2)$&	$(4,	6,	3,	5)$&7.524&0.398\\
16&		4/12/2024 8:38	&	&4/12/2024 8:50	&		753&		4/12/2024 8:50&		anonymous	&	4&	 $(2,1)$&		$(4,	6,	5,	3)$&	6.624&	0.603\\
17&	4/12/2024 8:25	&&4/12/2024 8:51	&	1576	&4/12/2024 8:51&	anonymous	&3& 	$(1,2)$	&$(4,	5,	3,	6)$&7.362&	0.407\\
18&	4/12/2024 8:42&	&4/12/2024 8:52&		633	&4/12/2024 8:52&	anonymous	&5&	$(2,1)$	&$(6,4,3,5)$	&	6.450&	0.775\\
19&	4/12/2024 8:53&	&4/12/2024 8:56	&	175	&4/12/2024 8:56&	anonymous	&5	&$(1,2)$	&$(4,3,5,6)$ &5.164&0.968\\
20&	4/12/2024 8:50&	&4/12/2024 9:03	&	753	&4/12/2024 9:03&	anonymous&	3	&$(1,2)$&	$(3,4,5,6)$&6.624&0.452\\
21&	4/12/2024 8:57&	&4/12/2024 9:06&		550&	4/12/2024 9:06&	anonymous&	2	&	$(1,2)$&	$(4,	6,	3,	5)$	&6.309&	0.316\\
22&4/12/2024 9:05&	&4/12/2024 9:08	&	179&	4/12/2024 9:08	&anonymous	&3	&	$(2,1)$	&$(4,	5,	3,	6)$	&5.187&	0.578\\
23&	4/12/2024 9:00&&	4/12/2024 9:08	&	514&	4/12/2024 9:08	&anonymous	&4	&	$(1,2)$	&$(3,4,6,5)$	&	6.242&	0.640\\
24&	4/12/2024 9:12	&&4/12/2024 9:22		&621&	4/12/2024 9:23&	anonymous	&5	&	$(1,2)$	&$(3,	6,	4,	5)$&6.431&	0.777\\
25&	4/12/2024 9:28&	&4/12/2024 9:39&		628&	4/12/2024 9:39&	anonymous	&6&	 	$(1,2)$	&$(6,3,	4,	5)$	&6.442&	0.931\\
26&	4/12/2024 9:33	&&4/12/2024 9:41	&	484	&4/12/2024 9:41&	anonymous	&4	&	$(1,2)$	&$(4,3,6,5)$&6.182&	0.647\\
27&4/12/2024 9:33&	&4/12/2024 9:54	&	1249&	4/12/2024 9:54&	anonymous	&4	&	$(1,2)$&$(4,	5,	3,	6)$&7.130&	0.561\\
28&	4/12/2024 9:57&&	4/12/2024 10:01&		272&	4/12/2024 10:01	&anonymous	&0	&$(2,1)$	&$(4,	6,	3,	5)$&5.605&	0\\
29&	4/12/2024 10:06	&&4/12/2024 10:11	&	301&	4/12/2024 10:11&	anonymous	&5	&	$(1,2)$&$(4,3,6,5)$&	5.707&0.876\\
30&	4/12/2024 10:01&&	4/12/2024 10:15	&	887	&4/12/2024 10:15&	anonymous&	5	& 	$(1,2)$	&$(3,	6,	4,	5)$	&6.787&	0.736\\
31&	4/12/2024 10:21&&	4/12/2024 10:25	&	265	&4/12/2024 10:25&	anonymous	&2	& 	$(2,1)$	&$(4,3,5,6)$	&5.579&0.358\\
32&	4/12/2024 9:09	&&4/12/2024 10:30&		4833&	4/12/2024 10:30&	anonymous	&4	&	$(1,2)$ &$(3,4,6,5)$&8.483&	0.471\\
33&	4/12/2024 10:26	&&4/12/2024 10:37&		633	&4/12/2024 10:37&	anonymous	&3&	$(2,1)$	&$(6,4,3,5)$	&	6.450&0.465\\
34&	4/12/2024 10:37	&&4/12/2024 10:38&		60	&4/12/2024 10:38&	anonymous&	4	& $(2,1)$&	$(6,4,5,3)$	&4.094&	0.976\\
35&	4/12/2024 9:55&	&4/12/2024 10:53	&	3468	&4/12/2024 10:53&	anonymous	&3	& $(1,2)$&$(4,	6,	5,	3)$	&8.151&	0.368\\
36&	4/12/2024 10:57&&	4/12/2024 11:03&		360	&4/12/2024 11:03&	anonymous&	1	& 	$(1,2)$&$(4,	5,	3,	6)$&5.886&	0.169\\
37&	4/12/2024 11:09	&&4/12/2024 11:21&		756	&4/12/2024 11:21&	anonymous&	3	& 	$(2,1)$	&$(6,3,	4,	5)$&6.628&	0.452\\
38&	4/12/2024 11:27&&	4/12/2024 11:47	&	1175&	4/12/2024 11:47&	anonymous&	3&$(2,1)$	&$(4,	5,	6,	3)$	&7.069&	0.424\\
39&	4/12/2024 9:47&	&4/12/2024 11:53&	7541&4/12/2024 11:53&	anonymous&	4	& $(2,1)$&$(3,	6,	4,	5)$&	8.928&	0.448\\
40&	4/12/2024 12:01&&	4/12/2024 12:05	&	257	&4/12/2024 12:05&	anonymous	&3& $(2,1)$&$(4,3,6,5)$	&5.549&0.540\\
41&	4/12/2024 13:54&&	4/12/2024 13:59	&	319	&4/12/2024 13:59&	anonymous&	2	&	$(1,2)$	&$(6,4,3,5)$	&5.765&	0.346\\
42&	4/12/2024 18:39	&&4/12/2024 19:00	&	1286	&4/12/2024 19:00	&anonymous	&2&	$(2,1)$	&$(3,4,5,6)$&7.159&	0.279\\
43&4/12/2024 9:24&	&4/12/2024 20:15	&	39071&	4/12/2024 20:15&	anonymous&	2&		$(1,2)$&	$(4,	6,	5,	3)$&10.573&	0.189\\
44&	4/12/2024 20:37&	&4/12/2024 20:57	&	1187	&4/12/2024 20:57&	anonymous	&3	& 	$(1,2)$&	$(3,4,6,5)$	&	7.079&	0.423\\
45&	4/13/2024 5:25&	&4/13/2024 5:31	&	376&	4/13/2024 5:31	&anonymous&	4	& $(2,1)$	&$(4,3,5,6)$&5.929&0.674\\
46&	4/13/2024 5:25&	&4/13/2024 6:34	&	4110	&4/13/2024 6:34	&anonymous&	3	& 	$(1,2)$	&$(6,4,3,5)$	&8.321&	0.360\\
47&	4/13/2024 13:06	&&4/13/2024 13:07&		73	&4/13/2024 13:07&	anonymous	&0	&	$(1,2)$	&$(4,	5,	3,	6)$	&4.290&	0\\
48&	4/13/2024 14:36	&&4/13/2024 14:47	&	697	&4/13/2024 14:47	&anonymous&	2	&$(1,2)$	&$(3,4,6,5)$&	6.546&0.305\\
49	&4/12/2024 8:10	&&4/13/2024 16:21&		115859	&4/13/2024 16:21	&anonymous&	3	&$(2,1)$ &$(3,4,5,6)$	&11.660&	0.257\\
50 &4/13/2024 19:45&&	4/13/2024 19:51	&	396	&4/13/2024 19:51&	anonymous&	4	&$(2,1)$	&$(4,3,6,5)$&5.981&	0.668\\
51	&4/15/2024 7:40&&	4/15/2024 7:46	&	400&	4/15/2024 7:46&	anonymous	&4&		$(1,2)$&$(6,3,	4,	5)$	&5.991&	0.667\\
52&	4/15/2024 7:55&	&4/15/2024 8:07	&	729	&4/15/2024 8:07&	anonymous	&4	& $(2,1)$&$(3,	6,	4,	5)$&6.591&	0.606\\
53&	4/15/2024 11:29	&&4/15/2024 11:34	&	331	&4/15/2024 11:34&	anonymous	&3	& 	$(1,2)$&	$(4,	6,	3,	5)$&5.802&0.517\\
54&	4/16/2024 14:24	&&4/16/2024 14:28&		257&	4/16/2024 14:28&	qr&	2	& $(2,1)$	&$(3,4,5,6)$	&5.549&	0.360\\
55&	4/16/2024 14:27	&&4/16/2024 14:31&		229&	4/16/2024 14:31	&qr	&2	&	$(1,2)$	&$(4,3,6,5)$&	5.433&0.368\\
56&	4/16/2024 14:24	&&4/16/2024 14:32&		501	&4/16/2024 14:32	&qr&	0	& 	$(2,1)$&	$(6,4,5,3)$	&6.216&	0\\
57&	4/16/2024 14:28	&&4/16/2024 14:33	&	292	&4/16/2024 14:33&	qr	&0	& 	$(2,1)$&$(3,4,6,5)$&	5.676&	0\\
58&	4/16/2024 14:25&	&4/16/2024 14:33	&	502&	4/16/2024 14:33	&qr	&3	& 	$(1,2)$&$(3,	6,	4,	5)$&	6.218&	0.482\\
59&	4/16/2024 14:24	&&4/16/2024 14:35	&	669	&4/16/2024 14:35&	anonymous&	2	& 	$(1,2)$&	$(4,	6,	5,	3)$	&6.505&	0.307\\
60&	4/16/2024 14:31&&	4/16/2024 14:35	&	251	&4/16/2024 14:35&	qr	&3	& 	$(1,2)$	&$(3,4,6,5)$	&5.525&	0.542\\
61&	4/16/2024 14:31	&&4/16/2024 14:35&		262	&4/16/2024 14:35	&qr&	2	& $(2,1)$	&$(4,	5,	3,	6)$&5.568&	0.359\\
62&	4/16/2024 14:29	&&4/16/2024 14:35	&	362	&4/16/2024 14:35	&qr	&2	&$(2,1)$&$(6,3,	4,	5)$&5.891&	0.339\\
63&	4/16/2024 14:28	&&4/16/2024 14:36&		469	&4/16/2024 14:36&	qr&	4&$(1,2)$	&$(4,	6,	5,	3)$	&6.150&	0.650\\
64&4/16/2024 15:51&	&4/16/2024 15:59&		429	&4/16/2024 15:59&	qr	&2&	 $(2,1)$&	$(6,4,3,5)$	&6.061&	0.329\\
65&	4/16/2024 14:24&&4/16/2024 16:09		&6314&	4/16/2024 16:09	&qr&	3	&$(1,2)$	&$(4,3,5,6)$&8.750&0.342\\
66&	4/16/2024 14:35&&	4/16/2024 14:48&		780	&4/9/2024 13:38&	qr&	3&		$(1,2)$&	$(4,	5,	6,	3)$	&6.659&	0.450\\\hline\et}
\end{table}

\end{document}